\documentclass{article}
\usepackage{arxiv}
\usepackage[utf8]{inputenc} 
\usepackage[T1]{fontenc} 
\DeclareUnicodeCharacter{03A9}{\ensuremath{\Omega}}
\DeclareUnicodeCharacter{039B}{\ensuremath{\Omega}}
\usepackage{url}            
\usepackage{booktabs}       
\usepackage{amsfonts}       
\usepackage{nicefrac}	
\usepackage{graphicx}
\usepackage{natbib}

\usepackage{amsmath}
\usepackage{float}
\usepackage{amsmath, amsthm, amssymb, amsfonts}
\usepackage{thmtools}
\usepackage{graphicx}
\usepackage{setspace}
\usepackage{appendix}
\usepackage[english]{babel}
\usepackage{framed}
\usepackage[dvipsnames]{xcolor}
\usepackage{tcolorbox}
\usepackage{csquotes}
\usepackage{caption}
\usepackage{subcaption}
\usepackage{parskip}
\usepackage{hyperref}
\hypersetup{
  colorlinks   = true, 
  urlcolor     = blue, 
  linkcolor    = black,
  citecolor    = blue 
}
\title{Reconstructing FHDE with Scalar and Gauge Fields}

\author{ \href{https://orcid.org/0000-0002-2842-3334}{\includegraphics[scale=0.06]{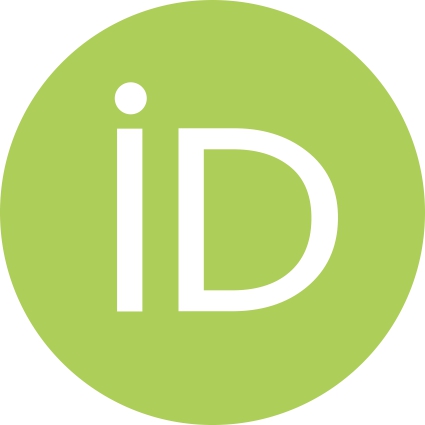}\hspace{1mm}Ayush Bidlan}\thanks{i21ph018@phy.svnit.ac.in} \\
	\textit{Department of Physics,}\\
	\textit{Sardar Vallabhbhai National Institute of Technology,}\\
	\textit{Surat 395007, Gujarat, India} \\
	\And
	\href{https://orcid.org/0000-0001-7170-8952}{\includegraphics[scale=0.06]{orcid.jpg}\hspace{1mm}Paulo Moniz}\thanks{pmoniz@ubi.pt} \\
	\textit{Departamento de Física, Centro de Matemática e Aplicações (CMA-UBI),}\\
	\textit{Universidade da Beira Interior,}\\
	\textit{Marquês d’Avila e Bolama, 6200-001 Covilhã, Portugal} \\
    \And
    \href{https://orcid.org/0000-0002-2086-4127}{\includegraphics[scale=0.06]{orcid.jpg}\hspace{1mm}Oem Trivedi}\thanks{oem.t@ahduni.edu.in}\\
    \textit{International Centre for Space and Cosmology,}\\
    \textit{Ahmedabad University,}\\
    \textit{Ahmedabad 380009, India}\\
}

\begin{document}
\maketitle

\begin{abstract}
	We revisit the Fractional Holographic Dark Energy (FHDE) model to reconstruct it by means of dynamic candidates such as ($i$) Quintessence, ($ii$) K-essence, ($iii$) Dilaton, ($iv$) Yang-Mills condensate, ($v$) DBI-essence, and ($vi$) Tachyonic fields in a flat Friedmann-Robertson-Walker (FRW) Universe. In particular, the dark-energy possibilities ($i$)-($vi$) are formulated through suitable field descriptions. Being concrete, we establish a comprehensive correspondence between FHDE and suitable scalar and gauge field frameworks that co-substantiate our investigation and subsequent discussion. In more detail, we methodically compute the corresponding  Equation of State (EoS) parameters and field (kinetic and potential)  features for the fractional parameter ($\alpha$) range, viz. $1<\alpha\leq2$. Conclusively, our results show that the modifications brought by the fractional features satisfactorily enable late-time cosmic acceleration, together with avoiding quantum instabilities by preventing the EoS from entering the phantom divide i.e., $\omega(z)\rightarrow-\infty$, which is a common issue in standard scalar field models without fractional dynamics (e.g., K-essence field). Our findings further indicate that fractional calculus attributes can be significant in addressing the challenges of dark-energy models by offering a robust framework to prospect late-time acceleration and properly fitting observational constraints. Notably, we find that as the fractional features start to dominate, the EoS parameter of all the effective field configurations asymptotically approaches a $\Lambda$CDM behaviour in the far-future limit $z\rightarrow-1$. In summary, the recent perspective introduced by FHDE \citep{Trivedi:2024inb} can indeed be cast as a promising aspirant through the use of prominent field frameworks.
\end{abstract}

\section{Introduction}
The mysterious nature of dark energy has been teasing us, constituting so far an obstacle blocking our
understanding of the Universe ever since its observational inception in the late 
twentieth century \citep{Riess_1998, Perlmutter_1999}. Observations of phenomena such as Type Ia Supernovae (SNe Ia) \citep{astier2006supernova, riess2004type, riess2007new, wood2007observational, davis2007scrutinizing, kowalski2008improved}, the Cosmic Microwave Background (CMB) \citep{spergel2003first,spergel2007three,komatsu2009five, Komatsu_2011} and Baryon Acoustic Oscillations (BAOs) \citep{eisenstein2005detection, percival2007measuring, percival2010baryon} have vindicated 
the presence of dark energy, estimated to constitute a substantial 
proportion ($\sim70\%)$ of the Universe. The traditional 
method for tackling late-time cosmic acceleration has been based on the addition of a cosmological constant, denoted by $\Lambda$, to the field equations of the General Theory of Relativity \citep{Weinberg:1988cp, Padmanabhan:2002ji}. This constant serves as a type of dark energy that permeates space, inducing a peculiar dynamics  
that opposes e.g. Newtonian attractive gravity,  fuelling an accelerated expansion of the Universe. The scenario, including the features previously outlined, constitutes our current understanding of the Universe on large scales and is
known as the $\Lambda$CDM model \citep{turner1997caselambdacdm, 1984Natur.311..517B}. 
It is pertinent to emphasize that the cosmological constant $\Lambda$ represents the simplest theoretical framework in order to explain the observational evidence of an accelerating Universe. 

The success of the $\Lambda$CDM scenario notwithstanding, 
it still struggles with the problem of how the particular value of 
$\Lambda$ is determined and finely tuned  \citep{Perivolaropoulos:2021jda, Condon:2018eqx}. 
In classical physics, a small value of the cosmological constant $\Lambda$ does not pose any major problems in theory; 
it is just another parameter in the theory. However, in Quantum Field Theory (QFT), 
vacuum fluctuations become crucial,  significantly impacting 
the overall (vacuum) energy density. These 
issues 
are at the origin of the challenge known as the "Old Cosmological Constant Problem" \citep{Lombriser:2019jia, Weinberg:1988cp}. Subsequently, a plethora of different models for dark energy were then proposed to explain the late-time acceleration of the Universe without the cosmological constant, constituting an attempt to fix the fine-tuning problem. 
They are, in good measure, phenomenological settings that include rolling fields in the presence of a suitable potential;  it generalises the cosmological constant through a concrete dynamic behaviour (NB. when the fields are not rolling, their potential energy $V\left(\varphi\right)$ behaves as a cosmological constant \citep{COPELAND_2006}). Let it be added that the main focus has been on spin-zero fields. 
Moreover, research was also conducted so as to 
generalise this strategy, 
employing higher spin dark energy models, such as spinors \citep{PhysRevD.72.123502, Cai_2008, yajnik2011darkenergyferromagneticcondensation, tsyba2011reconstructionfessencefermionicchaplygin}, vectors \citep{Armend_riz_Pic_n_1999, Zhao_2006}, and the $p-$form field \citep{Koivisto_2009, gupta2011darkenergychernsimonslike}.  Furthermore,
alternative explanations for the Universe's late-time acceleration were also constructed, proposing gravitational frameworks 
beyond General Relativity. E.g., considering modified gravity theories and string-theoretic approaches from quantum gravity \citep{Nojiri:2003jn, Nojiri:2003ft,Nojiri:2000kz,Nojiri:2003vn,Nojiri:2004ip,Nojiri:2004pf,Nojiri:2005pu,Nojiri:2005sr,Nojiri:2005sx,Nojiri:2006gh}. 
Based on the latter proposals, there are  current conjectural conditions, as suggested in \citep{cardone2004unified, paliathanasis2021dynamics, garg2019bounds}, 
that have been advocating
the construction of new cosmological models with more than one scalar field, 
aiming at an
effective field set-up 
to be 
consistent with quantum gravity \citep{bravo2020tip,achucarro2019string}. 
Furthermore, concerning information from observational data, it is important to affirm that we do not have unequivocal confirmation of 
any dynamics beyond the $\Lambda$CDM model. However, recent observational data by DESI Collaboration \citep{desicollaboration2024desi2024vicosmological, lodha2024desi2024constraintsphysicsfocused, Calderon_2024} hints at possible deviations,  prompting the exploration of models where $\Lambda$ may not be constant, in an attempt to stretch our outlook beyond the standard model of cosmology. Accordingly, a distinctive and vast literature has explored the dark energy issue, but questions regarding its subtleties remain strong.

 Within the broad context that the above paragraphs convey, our specific mindset is Holographic Dark Energy (HDE), proposed by Li \citep{Li:2004rb}, which stems from the association of an infrared (IR) cut-off of a quantum field theory (QFT) with the Holographic Principle (HP) \citep{Susskind:1994vu}. 
 In particular, we continue our research programme
promoting and further strengthening an approach 
that bridges the HP and elements from Fractional Calculus \citep{FC, Miller1993AnIT, Grigoletto2013FractionalVO};
it was designated as Fractional Holographic Dark Energy (FHDE) and was generically elaborated in \citep{Trivedi:2024inb}, supported with concrete examples of applications. 
Our main objective in this paper is, therefore, to make progress in investigating whether FHDE  can be a realistic and valued contender to explain the late-time accelerated expansion of the universe, in particular, by means of providing concrete field frameworks that can satisfactorily encompass the alluring dynamic properties of FHDE. 
 Within this strategy, we will import  features from 
 scalar and 
 gauge fields, suitably assembled into dark energy models while considering the holographic vacuum energy. In more detail, we employ 
 several phenomenological models such as ($i$) Quintessence, ($ii$) Kinetic Quintessence (K-essence), ($iii$) Dilaton, ($iv$)Yang-Mills condensate, ($v$) Dirac-Born-Infeld-essence (DBI-essence), and ($vi$) Tachyon scalar field as favourable inductors of dark energy dynamics.
As present in the literature, a considerable number of those models 
are characterized by having 
rolling fields because in this way 
a generalisation of the cosmological constant can be made, by means now of a 
dynamic behaviour of dark energy i.e., when the fields are not rolling, their potential energy $V\left(\varphi\right)$ behaves as a cosmological constant \citep{COPELAND_2006}. Let us remark again  that currently, we do not have the observational data needed to ascertain any dynamics beyond $\Lambda$CDM. However, certain data \citep{desicollaboration2024desi2024vicosmological, lodha2024desi2024constraintsphysicsfocused, Calderon_2024}
has hinted at possible deviations, prompting the exploration of models in an attempt to stretch our arm beyond the standard model of cosmology.
%


In our herewith concrete analysis aligned with FHDE, we
will consider 
homogeneous 
effective field configurations as candidates of dynamical dark energy (cf. ($i$)-($vi$) in the above paragraph), studying their evolution 
as a function of redshift (or time): this offers the flexibility to probe the late-time cosmic expansion beyond the restrictions of the  $\Lambda$CDM model 
by focusing on effective field configurations. Our approach allows for the exploration of a broader set of
dark energy models that specifically
account for the time-dependent evolution of, for example, the Equation of State (EoS) parameter $\omega(z)$, beyond the static framework of $\Lambda$CDM model ($\omega_{\Lambda\text{CDM}}=-1$).  
We, therefore, retrieve  an  EoS parameter by establishing a correspondence between the FHDE model and  effective field configurations as dark energy candidates, followed by presenting 
plots on the cosmological evolution of the 
field potential, $V\left(\varphi\right)$, and the 
field kinetic energy, $X$, over the redshift range $-1<z\leq2$. Such a holographic reconstruction has been explored in the recent literature,
considering several types of cut-off(s), employing different entropic corrections such as Tsallis and Barrows in modified gravity theories (see ref. \citep{Sheykhi_2011,ZHANG20071, WU2008152, SHEYKHI2010329, KARAMI201061,ualikhanova2024holographicreconstructionkessencemodel} and \citep{universe8120642} for further comments about this procedure). 
Throughout our paper, we take this idea one step further  supported by  the lens of the FHDE model 
\citep{Trivedi:2024inb}, in particular with the Hubble horizon as an IR cut-off by incorporating entropic correction motivated by fractional calculus (see \citep{Jalalzadeh_2021}). Our interest in this problem was majorly motivated by the work in \citep{Sheykhi_2011, WU2008152, Sheykhi_2010}.


We structure our work as follows. In section \ref{Sec-2}, we present the choice of IR (infrared) cut-off, within which we define the cosmological evolution of the different 
dark energy 
candidates ($i$)-($vi$). Subsequently, in section \ref{Sec 3}, we derive the EoS parameter along with constructions for the 
field potential $V\left(\varphi\right)$ and kinetic term $X$ for various dark energy models. Our results will then be 
conveyed through the assistance 
of an adequate
plots representing the cosmological evolution for each dark energy candidate over a redshift range of $-1<z\leq2$ for different values of the fractional parameter $\alpha$ in the range $1<\alpha\leq2$. We then proceed to section \ref{conc}, where we discuss our results and provide an outlook for future explorations within the paradigm of the FHDE model.


\section{Fractional HDE with Hubble horizon as its IR Cut-Off}\label{Sec-2}

 Before proceeding further, we must choose the cut-off scale $L$. As found in the literature, the initial suggestion consisted of taking a cut-off scale given by $L \to H^{-1}$, termed the Hubble horizon cut-off. This choice aimed to alleviate the fine-tuning problem by introducing a natural length scale associated with the inverse of the Hubble parameter,  $H$, but it was found that this particular scale resulted in the dark energy EoS parameter approaching zero, and it also failed to contribute significantly to the current accelerated expansion of the Universe \citep{Granda_2008}. An alternative idea that followed was then to consider 
instead of the particle horizon as the length scale:
\begin{equation} \label{lp}
    L_{\text{p}} = a \int_{0}^{t} \frac{dt}{a}. \quad
\end{equation}

Such an alternative yielded an EoS parameter higher than $-1/3$, but the challenges of explaining the present acceleration remained unresolved despite this modification. Another line of exploration included the future event horizon as the length scale:\begin{equation} \label{lf}
     L_{\text{f}} = a \int_{a}^{\infty} \frac{dt}{a}.
\end{equation}

Although the desired late-time acceleration regime can be achieved in this case, concerns were raised regarding causality \citep{Kim_2013}. Another option  was the Granda-Oliveros cut-off scale, which took into account the derivative of $H$ into the definition of $L$ \citep{Granda_2008}: \begin{equation}
    L_{\text{G-O}} = (\alpha H^2 + \beta \dot{H} )^{- \frac{1}{2}}.
\end{equation}

The most generalized proposal for a cut-off is the Nojiri-Odintsov cut-off (see \citep{Nojiri:2005pu, Nojiri:2017opc}, and \citep{nojiri2019holographic} for a detailed review), where we 
have the following form:\begin{equation} \label{nocut-off}
    L_{\text{N-O}} =  L(H,\dot{H},\ddot{H},...L_{\text{p}},L_{\text{f}}, \dot{L}_{\text{p}},\dot{L}_{\text{f}}...). 
\end{equation}

An additional issue with all these cut-off(s) relates to the classical stability of these models against perturbations \citep{Myung:2007pn}, which is present in all classes of HDE energy densities with various cut-off(s). In general, we  can have HDE models where the cut-off 
could be of the form of a functional as described in Eq. (\ref{nocut-off}), bearing  functions of $H$, $L_{\text{p}}$, and $L_{\text{f}}$ 
. Although all HDEs can be written in the form as described in Eq. (\ref{nocut-off}), 
it is difficult to motivate any of those cut-off(s) from common physical grounds. This leads to the formulation of various HDE theories being supported by different physical and mathematical considerations. In our work herein,  we would like to consider the Hubble horizon as an IR cut-off. Our reasoning for making this choice is two-fold. On the one hand, we are interested in the simplest way FHDE can produce universal 
evolution. By universal evolution, we mean that the FHDE model can produce a consistent late-time dark energy behaviour independent of specific assumptions about the cut-off scales, which makes it broadly applicable to various cosmological conditions. In that regard, the Hubble horizon cut-off provides a suitable choice
because it is directly related to the local expansion rate of the Universe and avoids issues of causality associated with the event horizon. Its simplicity and direct relevance to current cosmological conditions make it a natural candidate for the FHDE model. On the other hand, 
the Hubble horizon cut-off has been shown 
not to work within various HDE frameworks \citep{Li:2004rb, Myung:2007pn}
and so if FHDE can provide an observational consistent scenario 
then, it would represent a step forward in the ongoing appraisal of the Hubble cut-off. With this in mind, the Hubble cut-off; $L \to H^{-1}$ gives: \begin{equation} \label{h1frarho}
 \rho_{\text{de}} = 3 c^2 H^{\frac{3 \alpha - 2}{\alpha}}. 
\end{equation} 

We also define the fractional density parameters for dark energy and dark matter as follows:
\begin{equation} \label{omegas}
    \Omega_{\text{de}} = \frac{\rho_{\text{de}}}{3 H^2} \to  c^2 H^{\frac{\alpha - 2 }{\alpha}};\quad \Omega_{\text{dm}} = \frac{\rho_{\text{dm}}}{3 H^2}.
\end{equation}

This implies that the Friedman equation takes the following form:
\begin{equation}\label{FE}
\Omega_{\text{de}}+\Omega_{\text{dm}}=1.
\end{equation}

The continuity equation for dark energy and dark matter takes the form:
\begin{equation}\label{C-de}
\dot{\rho}_{\text{de}}+3H\rho_{\text{de}}(1+\omega_{\text{de}})=0,
\end{equation}
\begin{equation}\label{C-dm}
    \dot{\rho}_{\text{dm}}+3H\rho_{\text{dm}}(1+\omega_{\text{dm}})=0.
\end{equation}

On algebraic manipulation of Eq. (\ref{FE}) and (\ref{C-dm}), and Eq. (\ref{C-de}) and Eq. (\ref{h1frarho}), we obtain the EoS parameter for the FHDE model as:
\begin{equation}\label{fhdeomega}
    \omega_{\text{de}}=-1+\frac{(3\alpha-2)(1-\Omega_{\text{de}})}{2\alpha-\Omega_{\text{de}}(3\alpha-2)}.
\end{equation}

We use this expression extensively to establish the correspondence between all the effective field configurations 
and the FHDE model. Note that in the limiting case, when the fractional features start to diminish as $\alpha \to 2$, we find that $\omega_{\text{de}}\rightarrow0$, which is what we usually observe when one considers the conventional HDE scenario: $\rho_{\text{de}}=3c^{2}H^{2}$; with the Hubble cut-off. The upshot of this discussion is that as the fractional features become more dominant, they play a key role in uncovering the dynamic behaviour of dark energy (see \citep{Trivedi:2024inb} for a detailed review) during late-time cosmic acceleration of the Universe. Appendix \ref{a} 
presents implicit expressions for some of the relevant cosmological parameters, like fractional dark energy density parameter, $\Omega_{\text{de}}(z)$, and Hubble parameter, $H(z)$,  which we will make use in our study as depicted in next section, namely with the assistance of several corroborating plots.  



\section{Reconstructing Fractional Holographic Dark Energy}\label{Sec 3}

In this section, we reconstruct the FHDE model for dark energy effective field candidates,  such as Quintessence, K-essence
, Dilaton, Yang-Mills condensate, Dirac-Born-Infeld-essence, and Tachyons. To be more specific, we  establish a correspondence between the FHDE and scalar field models in the following way:
\begin{equation}\label{correspondence}
    \omega_{i}\leftrightarrow \omega_{\text{de}},\hspace{2mm}\rho_{i}\leftrightarrow\rho_{\text{de}},\hspace{2mm}p_{i}\leftrightarrow p_{\text{de}},
\end{equation}
where, the subscript, $i$, acts as 
the dummy index representing our effective field configuration 
for establishing the correspondence.

\subsection{Quintessence}

The most popular alternative to $\Lambda$ has been the scalar field model of quintessence, which is dynamic, unlike the cosmological constant (see \citep{Steinhardt:2003st}). Several works, such as \citep{ZHANG20071} and \citep{WU2008152}, have presented an approach to reconstructing dynamical dark energy using a quintessence scalar field with holographic vacuum energy motivated by HP in quantum gravity. The quintessence field 
has a real-valued scalar field Lagrangian of the form \citep{Tsujikawa:2013fta, Zlatev:1998tr}:
\begin{equation}
    \mathcal{L}_{\text{q}}=X_{\text{q}}-V_{\text{q}}(\varphi),
\end{equation}
where $X_{\text{q}}\equiv\frac{1}{2}\partial_{\mu}\varphi\partial^{\mu}\varphi$ and $V_{\text{q}}(\varphi)$ is the field potential. In dynamic scalar field models as quintessence, the potential $V_{\text{q}}(\varphi)$ takes the role of $\Lambda$ and must possess some naturalness without fine-tuning. However, the potential receives quantum loop corrections through a high-energy physics perspective, making it lose its technical naturalness \citep{de_Putter_2007}.\footnote{An alternative approach is adopting non-canonical kinetic terms in the scalar field Lagrangian, leading to a model known as Kinetic Quintessence (K-essence), 
which we will discuss in more detail in the next subsection.} In this section, we present the correspondence between the FHDE model and the quintessence scalar field by using the 
Eq. (\ref{correspondence}). In this manner, we obtain the fractional holographic reconstructed expressions for the kinetic, $X_{\text{q}}$, and potential energy, $V_{\text{q}}(\varphi)$, of quintessence with the flat FRW metric. We begin with the action for a quintessence scalar field that reads as:
\begin{equation}
    \mathcal{S}_{\text{q}}=\int d^{4}x\sqrt{-g}\left[-\frac{1}{2}g^{\mu\nu}\partial_{\mu}\varphi\partial_{\nu}\varphi-V_{\text{q}}\left(\varphi\right)\right],
\end{equation}
where $g^{\mu\nu}$ is the inverse of the metric $g_{\mu\nu}=\text{diag}\left(-,+,+,+\right)$ such that the quintessence field has standard kinetic form. From here, we find the corresponding energy-stress-momentum tensor using the following expression:
\begin{equation}\label{E-M Tensor}
T_{\text{q}}^{\mu\nu}=\frac{\partial\mathcal{L_{\text{q}}}}{\partial\left(\partial_{\mu}\varphi\right)}\partial^{\nu}\varphi-g^{\mu\nu}\mathcal{L}_{\text{q}}.
\end{equation}

By using the Eq. (\ref{E-M Tensor}) for the quintessence scalar field Lagrangian, the pressure and energy density can be obtained as:
$$\rho_{\text{q}}(t)=\frac{1}{2}\Dot{\varphi}_{\text{q}}^{2}+V_{\text{q}}\left(\varphi\right) \quad\text{and}\quad
    p_{\text{q}}(t)=\frac{1}{2}\Dot{\varphi}_{\text{q}}^{2}-V_{\text{q}}\left(\varphi\right).$$

From this, and using the expressions for the energy and pressure densities as usual, the EoS parameter is found to be:
\begin{equation}\label{eos-q}
\omega_{\text{q}}=\frac{p_{\text{q}}}{\rho_{\text{q}}}=\frac{X_{\text{q}}-V_{\text{q}}\left(\varphi\right)}{X_{\text{q}}+V_{\text{q}}\left(\varphi\right)}.
\end{equation}
Here $X_{\text{q}}=\dot{\varphi}^{2}_{\text{q}}/2$. Comparing the FHDE EoS parameter (see Eq. (\ref{fhdeomega})) 
with the EoS parameter of the quintessence scalar field (see Eq. (\ref{eos-q})), $\omega_{\text{de}}=\omega_{\text{q}}$, we can subsequently write:
\begin{equation}
\frac{X_{\text{q}}-V_{\text{q}}\left(\varphi\right)}{X_{\text{q}}+V_{\text{q}}\left(\varphi\right)}=-1+\frac{(3\alpha-2)\left(1-\Omega_{\text{de}}\right)}{2\alpha-\Omega_{\text{de}}(3\alpha-2)}.
\end{equation}

Upon algebraic manipulation of the above-written equation, we obtain the following expressions for $X_{\text{q}}$ and scalar field potential $V_{\text{q}}(\varphi)$ as:
\begin{equation}\label{q-field}
X_{\text{q}}=\frac{\dot{\varphi}^{2}_{\text{q}}}{2}=\frac{3\Omega_{\text{de}}H^{2}\left(3\alpha-2\right)\left(1-\Omega_{\text{de}}\right)}{4\alpha-\Omega_{\text{de}}\left(6\alpha-4\right)},
\end{equation}

\begin{equation}\label{q-potential}
V_{\text{q}}\left(\varphi\right)=3H^{2}\Omega_{\text{de}}\left[1-\frac{\left(3\alpha-2\right)\left(1-\Omega_{\text{de}}\right)}{4\alpha-2\Omega_{\text{de}}\left(3\alpha-2\right)}\right].
\end{equation}

\begin{figure}[h!]
\centering
\begin{subfigure}{.5\textwidth}
  \centering
\includegraphics[width=0.9\linewidth]{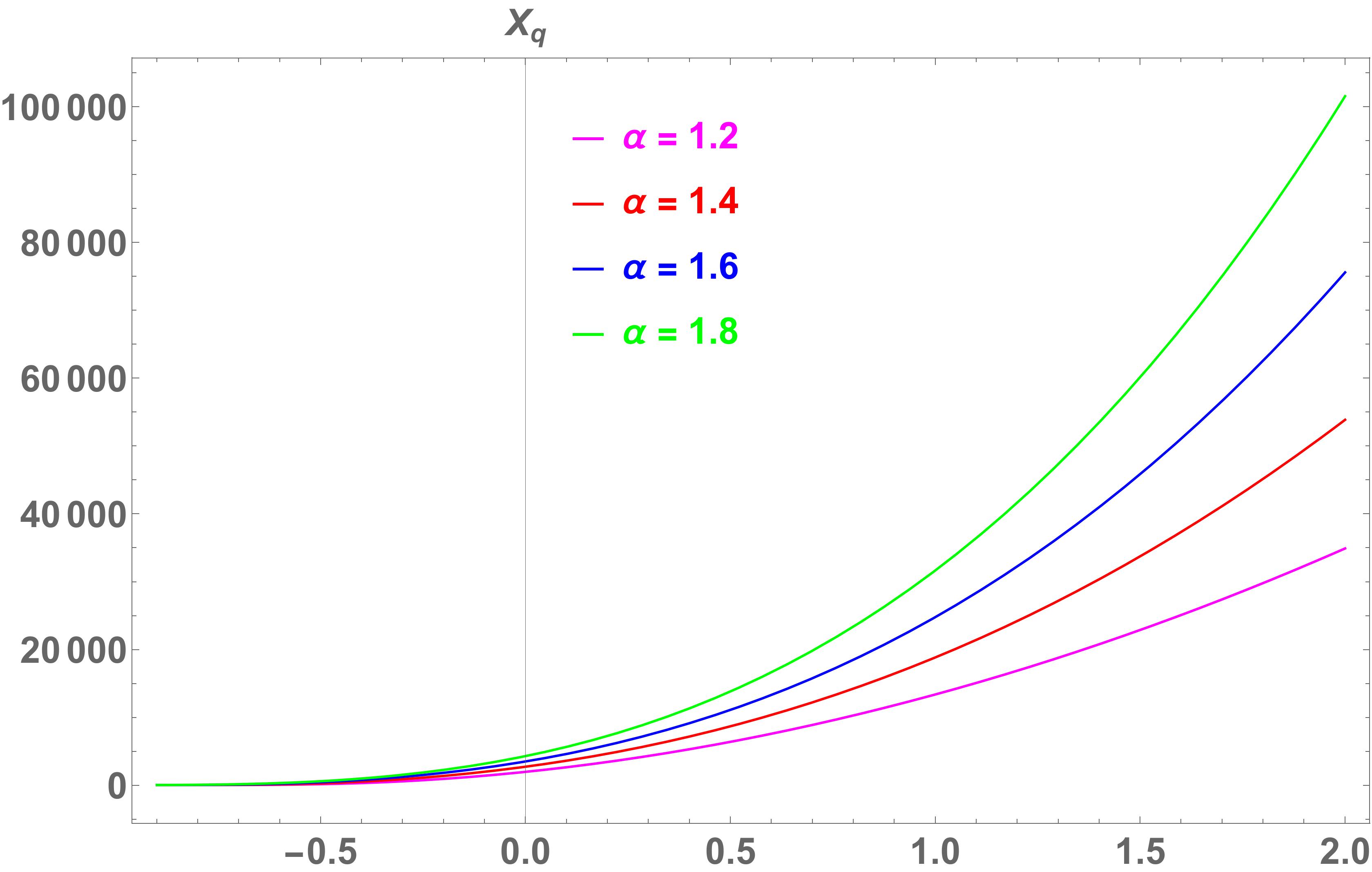}
  \caption{Plot of $X_{\text{q}}$ against redshift $z$.}
  \label{Figure 1: (a)}
\end{subfigure}%
\begin{subfigure}{.5\textwidth}
  \centering
  \includegraphics[width=0.9\linewidth]{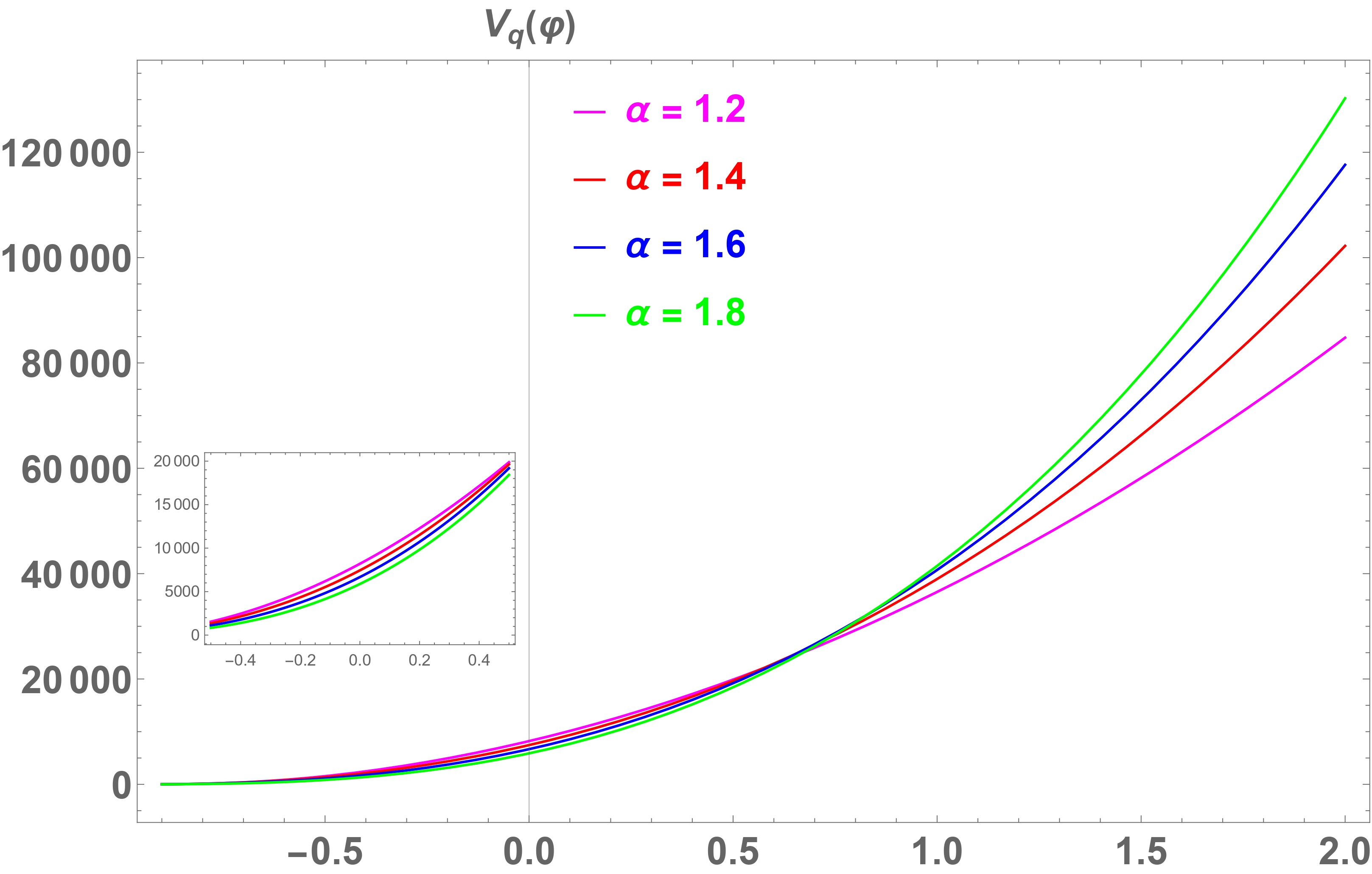}
  \caption{Plot of $V_{\text{q}}(\varphi)$ against redshift $z$.}
  \label{Figure 1: (b)}
\end{subfigure}
\caption{Plot for $X_{\text{q}}$ and $V_{\text{q}}(\varphi)$ against redshift $z$ for $\alpha=1.2, 1.4, 1.6$ and $1.8$.}
\label{Figure 1}
\end{figure}

Eq. (\ref{q-field}) and Eq. (\ref{q-potential}) describe the evolution of the kinetic energy, $X_{\text{q}}$, 
and potential energy, $V_{\text{q}}(\varphi)$,
of the quintessence field, respectively, with a varying redshift parameter $z$ in correspondence with the FHDE model. This evolution is illustrated in Figure \ref{Figure 1}.

\begin{figure}[ht]
\centering  \includegraphics[width=0.5\linewidth]{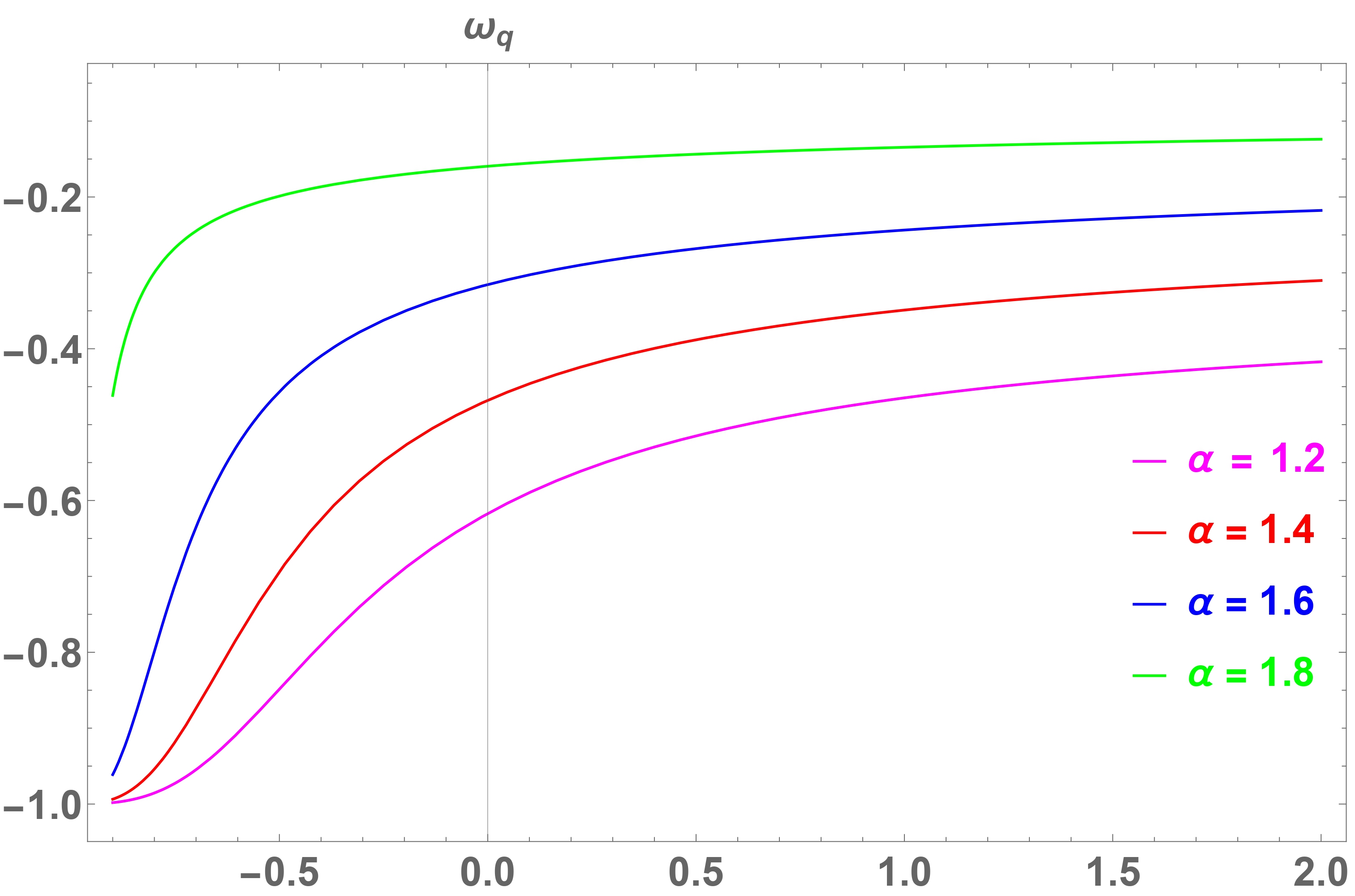}
\caption{Plot of $\omega_{\text{q}}$ against redshift $z$ for $\alpha=1.2, 1.4, 1.6$ and $1.8$; here $c=0.01$.}
\label{Figure 2}
\end{figure}


With detail:
\begin{itemize}
    \item In Figure \ref{Figure 1: (a)} and Figure \ref{Figure 1: (b)}, we can remark that $X_{\text{q}}$, and $V_{\text{q}}(\varphi)$, of the quintessence field in fractional holographic reconstruction, decays asymptotically, approaching zero as the redshift $z$ decreases into the far-future limit $z\rightarrow-1$. 

    In essence, let us then highlight the following: 

\begin{itemize}
    \item This behaviour is consistent across all values of the fractional parameter $\alpha$. 
    
    \item For larger values of $\alpha$ (e.g., $\alpha=1.8$ and $1.6$), $X_{\text{q}}$ and $V_{\text{q}}(\varphi)$ shows a highly dynamic evolution\footnote{By highly dynamic evolution, we mean that the evolution of the field changes more rapidly and the contrary for less dynamic evolution as we approach far-future limit $z\rightarrow-1$.}.
    
    \item In contrast, the evolution is less dynamic for smaller values of $\alpha$  (e.g., $\alpha=1.4$ and $1.2$). 
    
    \item However, as the redshift decreases below $z<0.55$ (see in Figure \ref{Figure 1: (b)}), subtle differences in the behaviour of the potential emerge, with $\alpha=1.8$ showing more appropriateness\footnote{The most appropriate model should ensure $V_{\text{q}}\gg X_{\text{q}}$ even at late-times. The potential reduces more gradually for $\alpha=1.8$, ensuring $\omega_{\text{q}}$ remains close to $-1$ for an extended period.} 
    due to its less dynamic evolution as the Universe transitions to a dark energy-dominated phase.  As $z\rightarrow-1$, the kinetic and potential energies tend to zero asymptotically,  approaching a $\Lambda$CDM behaviour. 

    \end{itemize}
    
    \item In Figure \ref{Figure 2}, we plotted the EoS parameter, $\omega_{\text{q}}$, of the fractional holographic reconstructed quintessence model. A suitable evolution of $w_{\text{q}}\rightarrow-1$ is possible to be retrieved: we identify that as fractional features start to dominate i.e., for small values of $\alpha$, like $\alpha=1.2$ and $1.4$,
    the EoS parameter for the quintessence scalar field asymptotically approaches 
    to the $\Lambda$CDM behaviour: $\omega_{\Lambda\text{CDM}}=-1$, 
    during the dark energy-dominated phase. As larger values of $\alpha$ are considered, the $\omega_{\text{q}}$ values deviate more from the 
    $\Lambda$CDM behaviour 
    We note that for $\alpha$ from $\alpha=1.2$ to 
    $1.8$, the evolution of $\omega_{\text{q}}$ remains in the quintessence regime $-1<\omega_{\text{q}}<-1/3$ during far-future limit $z\rightarrow-1$. 

\end{itemize}

Summarising for this subsection, Figure \ref{Figure 1: (a)} suggests that higher values of $\alpha$ may not be suitable because they do not ensure a slow evolving field, not matching
a nearly constant dark energy behaviour. 
On the other hand, in Figure \ref{Figure 1: (b)}, a higher value for $\alpha$ is preferable (over smaller values) because it induces a suitable situation for the potential for redshift region $-1<z<0.55$.

\subsection{K-essence}
A useful manner to consider
Kinetic quintessence, typically abbreviated as K-essence, where "K" stands for Kinetic, is a generalisation of the scalar field model with a standard kinetic term, i.e. common quintessence. Unlike in Quintessence, the K-essence scalar field framework avoids the requirement of a field potential, hence attempting to evade loop corrections in high-energy physics. K-essence scalar field has a Lagrangian of the form:
\begin{equation}\label{Lagrangian-kinetic}
\mathcal{L}_{\text{kq}}=f_{\text{kq}}(\varphi)F(X)-V(\varphi).
\end{equation}

From Eq. (\ref{Lagrangian-kinetic}), the Lagrangian for a scalar field with the standard kinetic term can be obtained by setting $f_{\text{kq}}(\varphi)=1$ and $F(X)=X$, i.e. quintessence. The primary aim of the k-essence model is to describe the inflationary evolution in the absence of a potential by employing a general class of non-standard kinetic terms for a scalar field $\varphi$. In \citep{Armend_riz_Pic_n_1999}, the authors consider a general kinetic Lagrangian $P(\varphi, X_{\text{kq}})$ for K-essence models to have relevance to a large class of models, which consists of the scalar field $\varphi$ and $X_{\text{kq}}=-\Dot{\varphi}_{\text{kq}}^{2}/2$. The action for K-essence can be written as:
\begin{equation}\label{action-K}
\mathcal{S}_{\text{kq}}=\int d^{4}x\sqrt{-g}P\left(\varphi, X_{\text{kq}}\right).
\end{equation}

 Its corresponding energy-stress-momentum tensor provides us with the energy density $\rho_{\text{kq}}\left(\varphi, X_{\text{kq}}\right)$ and pressure $p_{\text{kq}}\left(\varphi, X_{\text{kq}}\right)$, which can be expressed as shown in \citep{Armend_riz_Pic_n_1999, Copeland:2006wr}:
$$\rho_{\text{kq}}\left(\varphi,X_{\text{kq}}\right)=f_{\text{kq}}\left(\varphi\right)\left(-X_{\text{kq}}+3X_{\text{kq}}^{2}\right)\quad \text{and}\quad p_{\text{kq}}\left(\varphi,X_{\text{kq}}\right)=f_{\text{kq}}\left(\varphi\right)\left(-X_{\text{kq}}+X_{\text{kq}}^{2}\right).$$

Here, the function $f_{\text{kq}}(\varphi)$ represents the nature of coupling with the scalar field (for a detailed review, see \citep{Armend_riz_Pic_n_1999}). Moreover, we calculate its EoS parameter $\omega_{\text{kq}}=p_{\text{kq}}/\rho_{\text{kq}}$.  We obtain:
\begin{equation}\label{EoS-K}
\omega_{\text{kq}}=\frac{X_{\text{kq}}-1}{3X_{\text{kq}}-1}.
\end{equation}

Comparing the FHDE EoS parameter (see Eq. (\ref{fhdeomega})) with the EoS parameter of the k-essence scalar field (see Eq.
(\ref{EoS-K})). We obtain:
\begin{equation}
\frac{X_{\text{kq}}-1}{3X_{\text{kq}}-1}=-1+\frac{(3\alpha-2)\left(1-\Omega_{\text{de}}\right)}{2\alpha-\Omega_{\text{de}}(3\alpha-2)}
\end{equation}

Upon algebraic manipulation of the above-written equation, we obtain the following expressions for  $X_{\text{kq}}$ and $f_{\text{kq}}(\varphi)$ as:
\begin{equation}\label{kq-field}
X_{\text{kq}}=-\frac{\Dot{\varphi}^{2}_{\text{kq}}}{2}=1-\frac{2(\alpha-2)}{\alpha(1+3\Omega_{\text{de}})-2(\Omega_{\text{de}}+3)},
\end{equation}
\begin{equation}\label{kq-p}
f_{\text{kq}}\left(\varphi\right)=\frac{3H^{2}\Omega_{\text{de}}(\alpha+3\alpha\Omega_{\text{de}}-2(3+\Omega_{\text{de}}))^{2}}{2(2+\alpha+2\Omega_{\text{de}}-3\alpha\Omega_{\text{de}})(\alpha(3\Omega_{\text{de}}-2)-2\Omega_{\text{de}})}.
\end{equation}

Eq. (\ref{kq-field}) and Eq. (\ref{kq-p}) describe the late-time cosmic evolution of the kinetic energy, $X_{\text{kq}}$,
and coupling function, $f_{\text{kq}}$, respectively, with a varying redshift parameter $z$ in correspondence with the FHDE model. This evolution is illustrated in Figure \ref{Figure 2}.

\begin{figure}[H]
\centering
\begin{subfigure}{.5\textwidth}
  \centering
\includegraphics[width=0.9\linewidth]{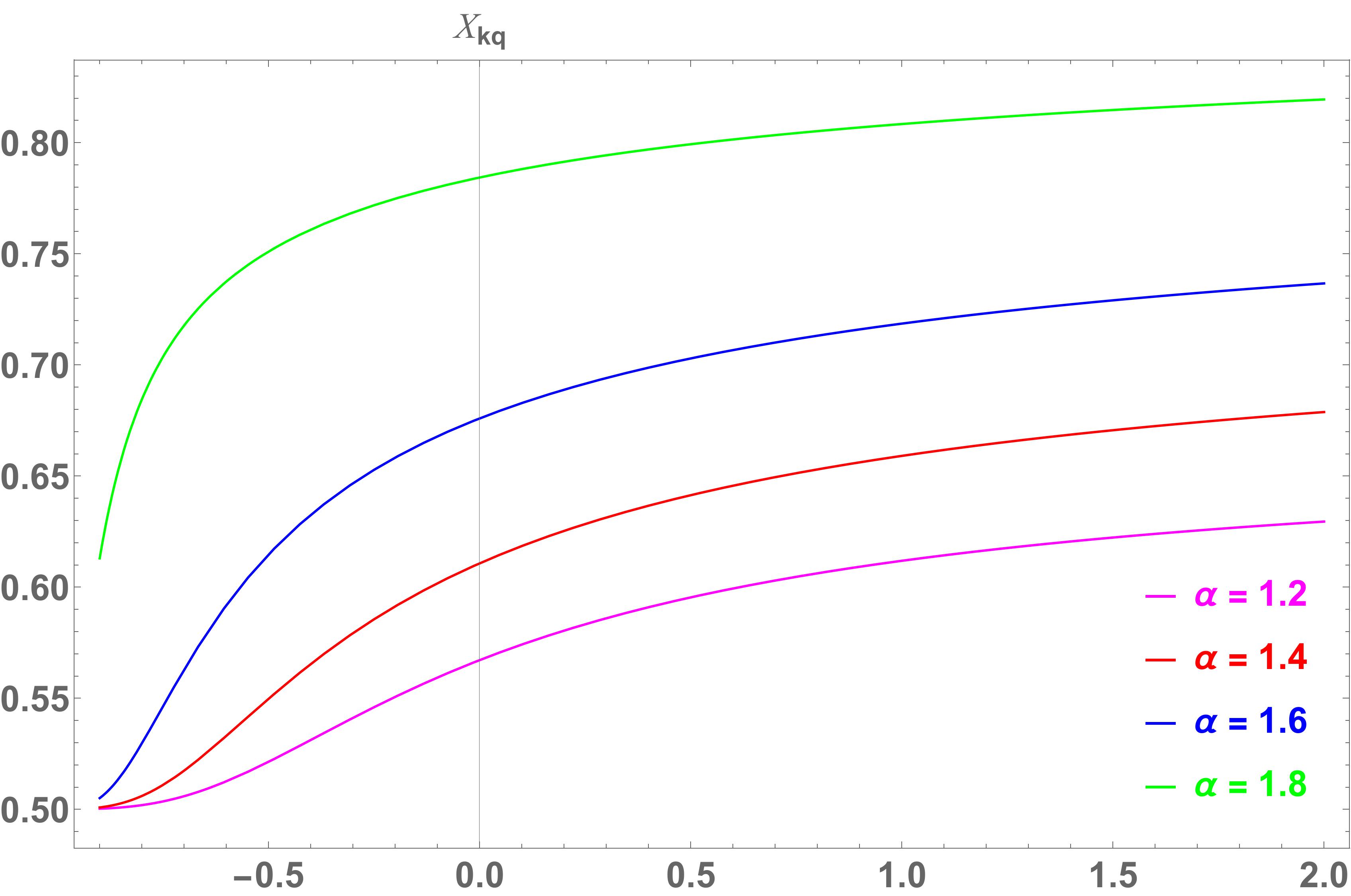}
  \caption{Plot of $X_{\text{kq}}$ against redshift $z$.}
  \label{Figure 3: (a)}
\end{subfigure}%
\begin{subfigure}{.5\textwidth}
  \centering
\includegraphics[width=1.0\linewidth]{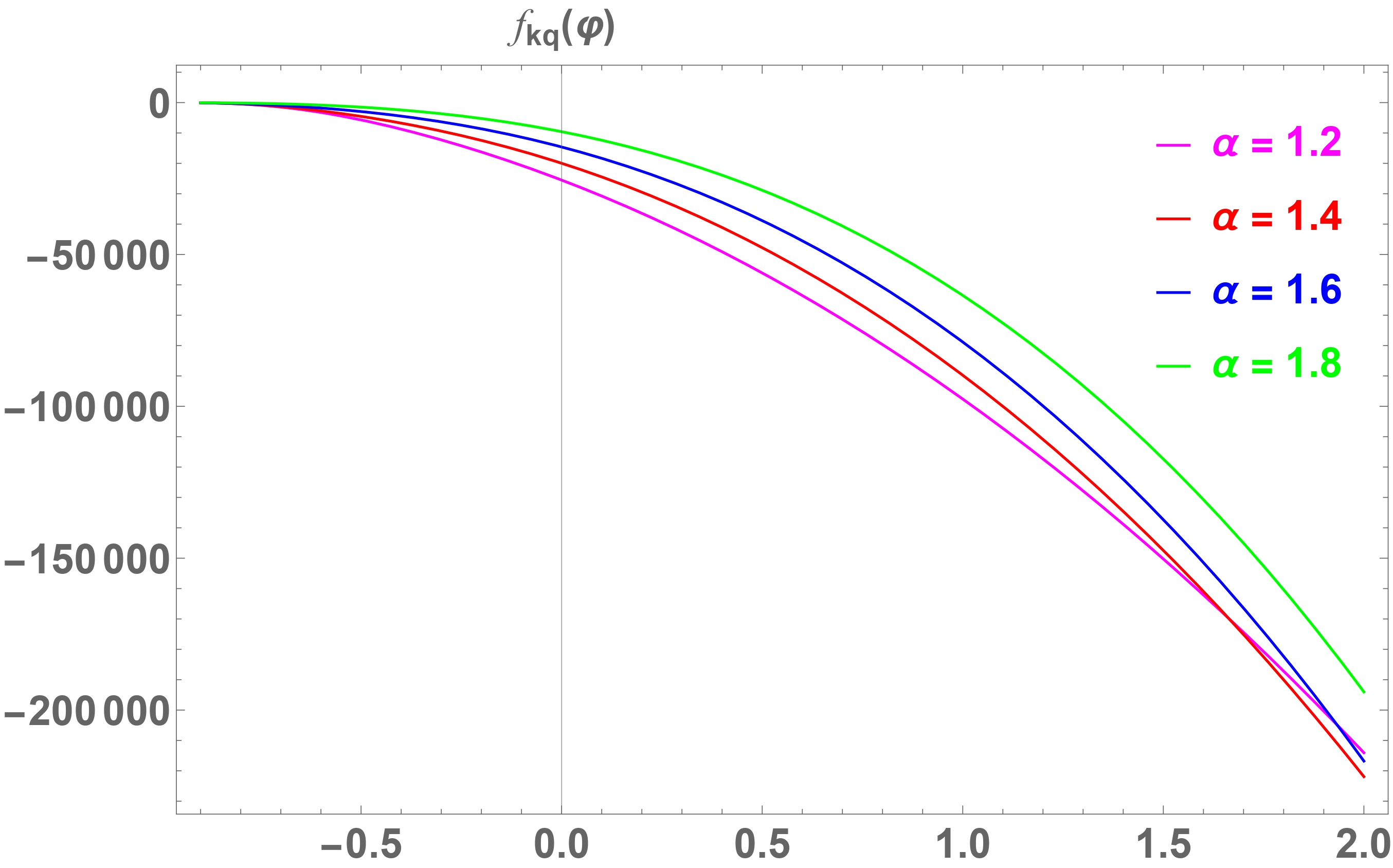}
  \caption{Plot of $f_{\text{kq}}(\varphi)$ against redshift $z$.}
  \label{Figure 3: (b)}
\end{subfigure}
\caption{Plot for $X_{\text{kq}}$ and $f_{\text{kq}}(\varphi)$ against redshift $z$ for $\alpha=1.2, 1.4, 1.6$ and $1.8$.}
\label{Figure 3}
\end{figure}
\begin{figure}[H]
\centering  \includegraphics[width=0.5\linewidth]{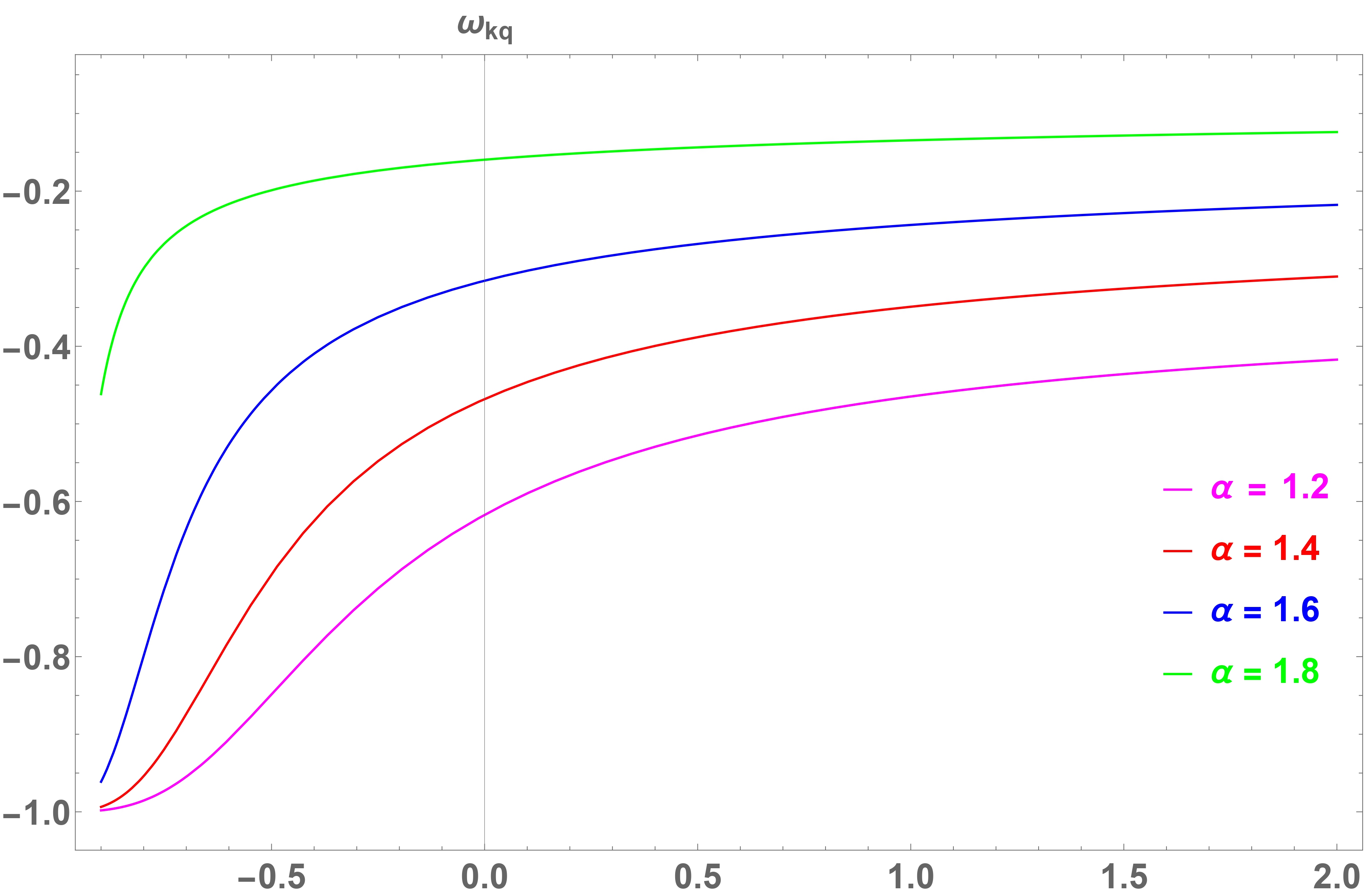}
\caption{Plot of $\omega_{\text{kq}}$ against redshift $z$ for $\alpha=1.2, 1.4, 1.6$ and $1.8$; here $c=0.01$.}
\label{Figure 4}
\end{figure}

We notice the following:
\begin{itemize}
    \item In Figure \ref{Figure 3: (a)} and Figure \ref{Figure 3: (b)}, we can observe that the kinetic energy, $X_{\text{kq}}$, and the coupling function, $f_{\text{kq}}(\varphi)$, of the K-essence field in fractional holographic reconstruction, decays asymptotically, approaching $X_{\text{kq}}\rightarrow1/2$, and $f_{\text{kq}}(\varphi)\rightarrow0$ as the redshift $z$ decreases in far-future limit $z\rightarrow-1$.

In particular: 

    \begin{itemize}

    \item For larger values of $\alpha$ (e.g., $\alpha=1.8$ and $1.6$), $X_{\text{kq}}$ shows a highly dynamic (or faster) evolution as redshift $z$ decreases. 
    
    \item In contrast, the evolution is less dynamic (or slower) for smaller values of $\alpha$  (e.g., $\alpha=1.4$ and $1.2$) as redshift $z$ decreases. 
    
    \item On the other hand, in Figure \ref{Figure 3: (b)}, we can remark that $f_{\text{kq}}(\varphi)$ increases from a smaller value at high redshifts to asymptotically approach zero at low redshifts. For larger values of $\alpha$, (e.g., $\alpha=1.8$ and $1.6$), $f_{\text{kq}}(\varphi)$ less dynamic (or slower) evolution is observed with decreasing redshift, whereas the evolution for smaller values of $\alpha$ (e.g., $\alpha=1.2$ and $1.4$), highly dynamic over the redshift region $-1<z\leq2$.

     \end{itemize}

    \item In Figure \ref{Figure 4}, we plotted for the EoS parameter, $\omega_{\text{kq}}$, of the fractional holographic reconstructed K-essence model. A suitable evolution of $\omega_{\text{kq}}\rightarrow-1$ is possible to be retrieved: we identify that as fractional features start to dominate i.e., for small values of $\alpha$, like $\alpha=1.2$ and $1.4$, the EoS parameter for K-essence scalar field asymptotically approaches to the $\Lambda$CDM behaviour, during dark energy dominated phase. We observe a similar behaviour for the EoS parameter for quintessence (see Figure \ref{Figure 2}) in the far-future limit.
\end{itemize}

To sum up, Figure \ref{Figure 3: (a)} suggests that higher values of $\alpha$ may not be suitable because they do not ensure a slow evolution of the field, as we remarked for Quintessence, in Figure \ref{Figure 3: (a)}. Therefore, higher values of $\alpha$ do not seem suitable,
to match a nearly constant dark energy behaviour.

 

\subsection{Dilaton}
Dilatonic Ghost condensate is a new type of dark energy model based on String Theory. In \citep{Piazza_2004}, the authors consider the classical dynamics of the system with a negative kinematic term $-X$ and the bell-type potential and estimate the result that the dilaton field evolves toward the potential maximum with an EoS: $\omega_{\text{d}}\leq-1$. However, since this system is unstable due to quantum fluctuations, a term of the type, $\exp{(\lambda\varphi)}X^{2}$, ensures stability at the quantum scale. When the conditions for the stability are satisfied, the EoS is restricted in the range from being greater than or equal to $-1$ in sharp contrast to the phantom field models of dark energy, as discussed in \citep{Carroll_2003}. The most general form of Lagrangian for a scalar field can be expressed as:
\begin{equation}
    \mathcal{L}_{\text{d}}=\frac{1}{2}(\partial\varphi)^{2}+\frac{A}{m^{4}}(\partial\varphi)^{4}\exp{\left(\frac{\lambda\varphi}{M_{P}}\right)} + \text{higher order terms}.
\end{equation}

Hereafter, we will set the Planck mass to unity and express the Lagrangian as:
\begin{equation}\label{pressure-D}
    p_{\text{d}}(X_{\text{d}},\varphi)=-X_{\text{d}}+\beta \exp{(\lambda\varphi)}X_{\text{d}}^{2},
\end{equation}
where $\beta=A/m^{4}$. The energy-stress-momentum tensor for this Lagrangian can be expressed as:
\begin{equation}
T_{\mu\nu}^{\varphi}=g_{\mu\nu}p_{\text{d}}+\frac{\partial p_{\text{d}}}{\partial X}\partial_{\mu}\varphi\partial_{\nu}\varphi.
\end{equation}

From this, the energy density can be obtained as:
\begin{equation}\label{dilaton-density}
    \rho_{\text{d}}(X,\varphi)=-X+3\beta \exp{(\lambda\varphi)}X^{2}
\end{equation}
where, $X_{\text{d}}=\dot{\varphi}^{2}/{2}$, and $\lambda$, and $\beta$ are constants. The EoS parameter for the dilaton model can be written as:
\begin{equation}\label{EoS-D}
\omega_{\text{d}}=\frac{-1+\beta\exp{(\lambda\varphi)}X_{\text{d}}}{-1+3\beta\exp{(\lambda\varphi)}X_{\text{d}}}.
\end{equation}

Now, we establish the correspondence between the FHDE EoS parameter and the Dilaton EoS parameter, and we obtain:
\begin{equation}
\frac{-1+\beta\exp{(\lambda\varphi)}X_{\text{d}}}{-1+3\beta\exp{(\lambda\varphi)}X_{\text{d}}}=-1+\frac{(3\alpha-2)\left(1-\Omega_{\text{de}}\right)}{2\alpha-\Omega_{\text{de}}(3\alpha-2)}
\end{equation}

Upon solving the above-written equation we obtain the expression for $X_{\text{d}}$, and $\beta\exp{(\lambda\varphi)X_{\text{d}}}$ as:

\begin{equation}\label{dilaton-kinetic}
X_{\text{d}}=\frac{\dot{\varphi}^{2}}{2}=\frac{3H^{2}\Omega_{\text{de}}(\alpha+3\alpha\Omega_{\text{de}}-2(3+\Omega_{\text{de}}))}{\alpha(6\Omega_{\text{de}}-4)-4\Omega_{\text{de}}},
\end{equation}
\begin{equation}\label{dilaton-exp}
    \beta\exp{(\lambda\varphi)}X_{\text{d}}=1-\frac{2(\alpha-2)}{\alpha+3\alpha\Omega_{\text{de}}-2(3+\Omega_{\text{de}})}.
\end{equation}

\begin{figure}[H]
\centering
\begin{subfigure}{.5\textwidth}
  \centering
\includegraphics[width=0.9\linewidth]{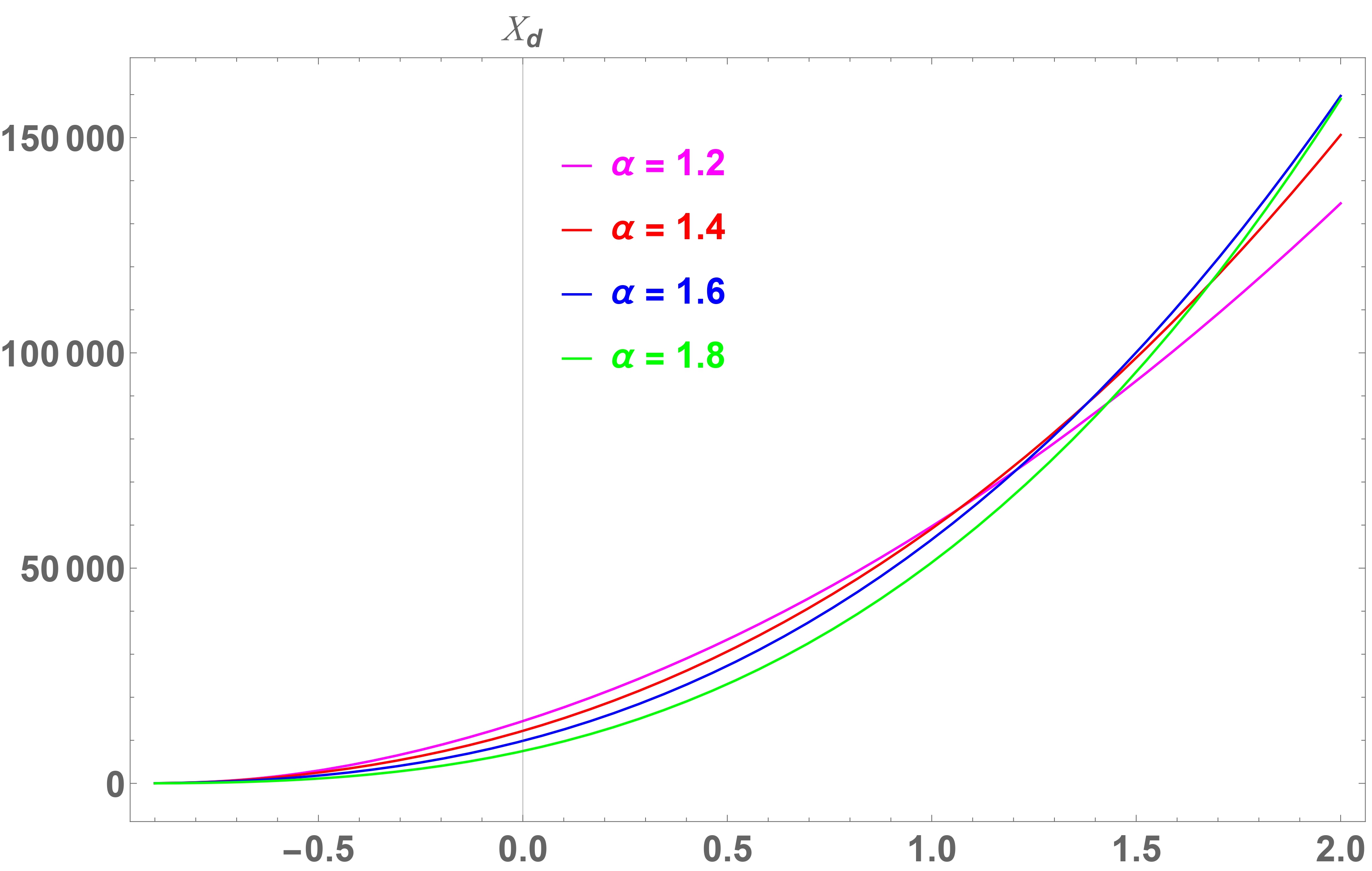}
  \caption{Plot of $X_{\text{d}}$ against redshift $z$.}
  \label{Figure 5: (a)}
\end{subfigure}%
\begin{subfigure}{.5\textwidth}
  \centering
  \includegraphics[width=0.9\linewidth]{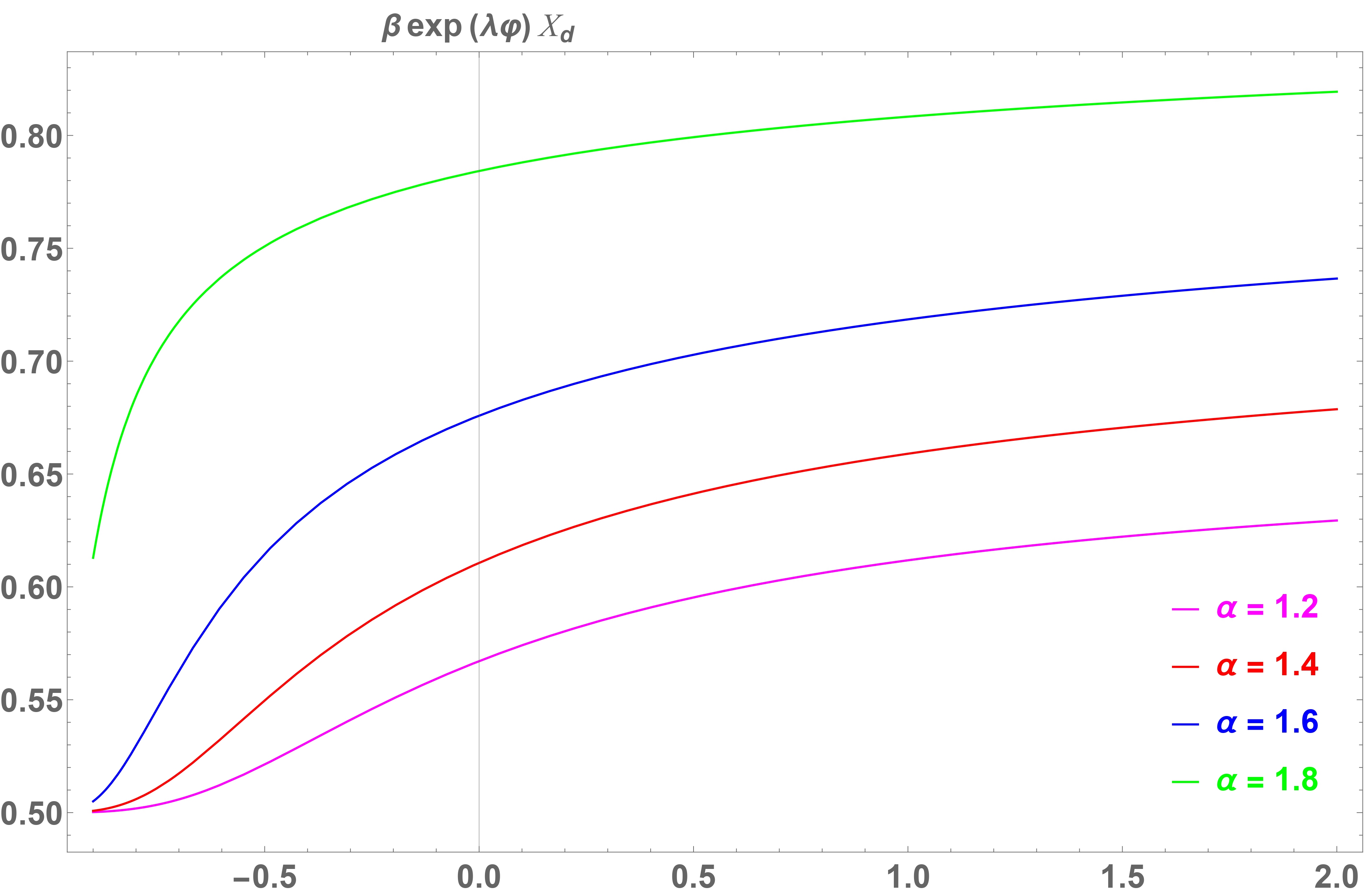}
  \caption{Plot of $\beta\exp{(\lambda\varphi)}X_{\text{d}}$ against redshift $z$.}
  \label{Figure 5: (b)}
\end{subfigure}
\caption{Plot of $X_{\text{d}}$ and $\beta\exp{(\lambda\varphi)}X_{\text{d}}$, respectively, against redshift $z$ for $\alpha=1.2, 1.4, 1.6$ and $1.8$.}
\label{Figure 5}
\end{figure}

Eq. (\ref{dilaton-kinetic}) and Eq. (\ref{dilaton-exp}) describe the late-time cosmic evolution of kinetic energy, $X_{\text{d}}$, and exponential potential, $\beta\exp{(\lambda\varphi)}X_{\text{d}}$, with varying redshift parameter $z$ in correspondence with the FHDE model. The evolution is illustrated in Figure \ref{Figure 5}. 

\begin{figure}[H]
\centering  \includegraphics[width=0.5\linewidth]{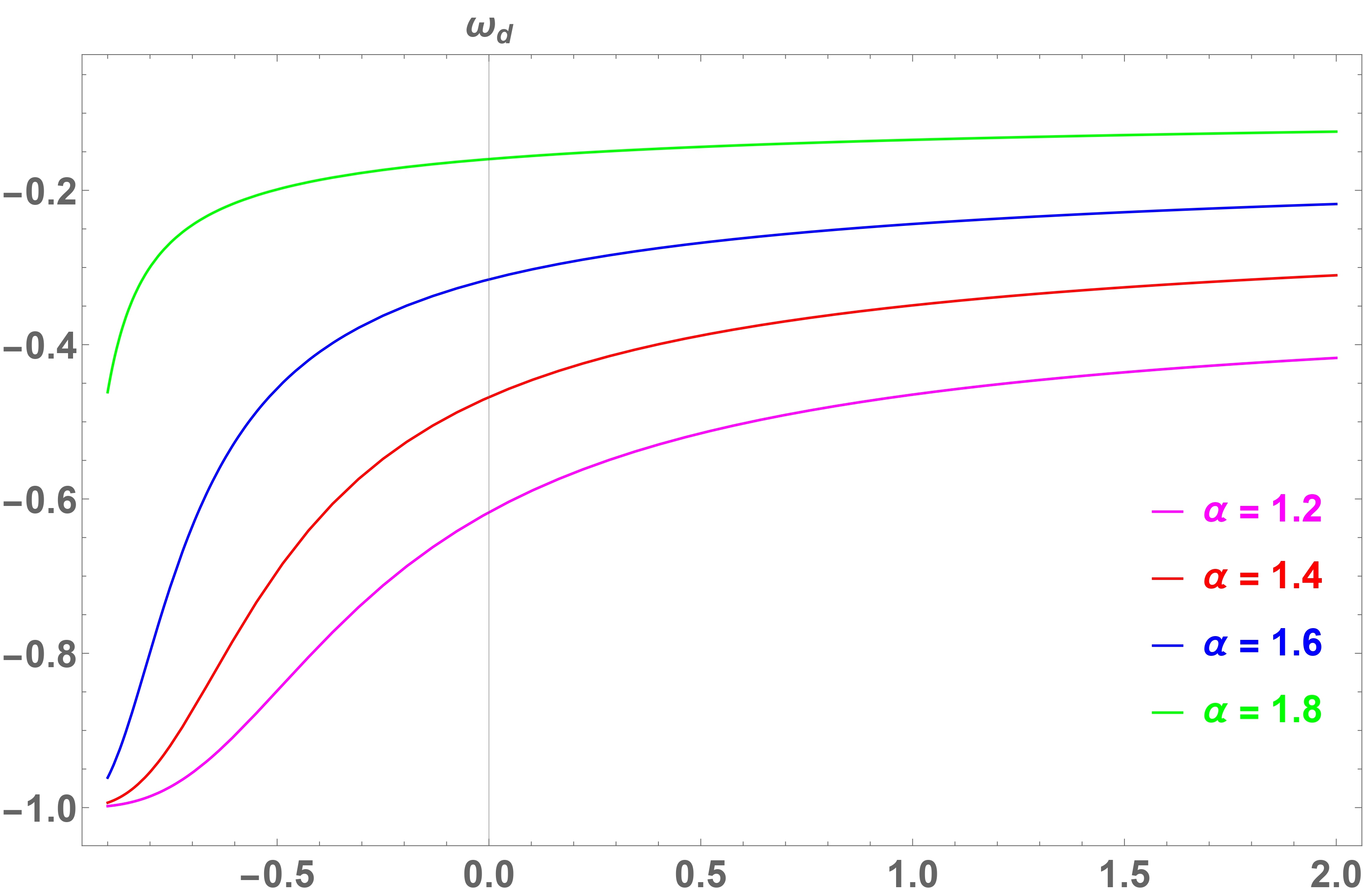}
\caption{Plot of $\omega_{\text{d}}$ against redshift $z$ for $\alpha=1.2, 1.4, 1.6$ and $1.8$; here $c=0.01$.}
\label{Figure 6}
\end{figure}

With detail:
\begin{itemize}
    \item In Figure \ref{Figure 5: (a)} and Figure \ref{Figure 5: (b)}, we observe that the kinetic energy, $X_{\text{d}}$, and the exponential function, $\beta\exp{(\lambda\varphi
)X_{\text{d}}}$, of the Dilaton field in fractional holographic reconstruction decay asymptotically, approaching $X_{\text{d}}\rightarrow0$, and $\beta\exp{(\lambda\varphi
)X_{\text{d}}}\rightarrow1/2$, respectively, as the redshift $z$ decreases in the far-future limit $z\rightarrow-1$.

In addition, we can notice the following:

\begin{itemize}
    \item For larger values of $\alpha$ (e.g., $\alpha=1.8$ and $1.6$), $X_{\text{d}}$ shows a highly dynamic (or faster) evolution, whereas the evolution is less dynamic (or slower) for smaller values of $\alpha$  (e.g., $\alpha=1.4$ and $1.2$) in the redshift region $1.5\leq z\leq 2$. 

\item As we take smaller and smaller values of redshift in the far-future limit $z\rightarrow-1$,  a less dynamic behaviour is perceived for large values of $\alpha$ and a highly dynamic behaviour for smaller values of $\alpha$. 

\item On the other hand, in Figure \ref{Figure 5: (b)}, 
we discern that for larger values of $\alpha$ (e.g., $\alpha=1.8$ and $1.6$), the potential evolution is highly dynamical (or faster) and in contrast, the evolution for smaller values of $\alpha$ (e.g., $\alpha=1.2$ and $1.4$), is less dynamic over the redshift region $-1<z\leq2$.

\end{itemize}

\item In Figure \ref{Figure 6}, we plotted for the EoS parameter, $\omega_{\text{d}}$, of the fractional holographic reconstructed Dilaton model. We can retrieve that a suitable evolution of $\omega_{\text{d}}\rightarrow-1$ is possible to identify as fractional features start to dominate because it represents a scenario where the K-essence scalar field behaves similarly to the $\Lambda$CDM model during dark energy dominated phase. 

We observe a similar behaviour for the EoS parameter for quintessence (see Figure \ref{Figure 2}) and K-essence (see Figure \ref{Figure 4}) in the far-future limit.
\end{itemize}

In a nutshell, Figure \ref{Figure 5: (a)} suggests that a higher value of $\alpha$ provides a suitable scenario because it ensures a slow-evolving field, making it 
suitable
to fit with a nearly constant dark energy behaviour. On the other hand, in Figure \ref{Figure 5: (b)}, a lower value of $\alpha$ is preferable (over larger values) because it induces a suitable situation\footnote{We retrieve that the exponential potential of Dilaton, $\beta\exp{(\lambda\varphi)X_{\text{d}}}$, behaves similarly to $X_{\text{kq}}$ (see Figure \ref{Figure 3: (a)}) and approaches a value of $1/2$ in the far-future limit $z\rightarrow-1$.} for the potential in the redshift region $-1<z\leq2$.

\subsection{Yang-Mills Field condensate}

The Yang-Mills Condensate (YMC) is an interesting approach to the problem of dark energy. It involves studying gauge boson fields in cosmological settings to investigate the cosmic evolution of components such as dark energy. The motivation for investigating YMC alongside the dynamic scalar field theories as a possible dark energy candidate is two-fold. First, the effective Yang-Mills Lagrangian is entirely determined by quantum field theory, which implies that the only parameter that can be changed is the energy scale of the theory. Thus, any modifications to the effective Lagrangian of the theory should be avoided. Second, in scalar field dark energy models such as quintessence, the EoS parameter lies in the range $-1<\omega_{\text{q}}<1$. In order to obtain $\omega_{\text{q}}<-1$, one needs to introduce a "phantom" field, which may bear quantum instabilities. However, that is avoided in the YMC dark energy model (see \citep{Zhao_2006} and \citep{zhao2009quantumyangmillscondensatedark} for a 
review). 


In an effective YMC dark energy model, the Lagrangian is given by \citep{Zhao_2006, PhysRevD.23.2905, ZHao1_2006} as,
\begin{equation}\label{Lag-YMC}
\mathcal{L}_{\text{YMC}}=\frac{b}{2}F\left(\ln|\frac{F}{\kappa^{2}}|-1\right),
\end{equation}
where $\kappa$ is the renormalisation scale of the dimension of squared mass, $b$ is Callan-Syamanzik coefficient $b=(11N-2N_{f})/24\pi^{2}$ for $SU(N)$ where $N_{f}$ represents the number of quark flavours, and $F\equiv-\frac{1}{2}F^{a}_{\mu\nu}F^{a\mu\nu}$ plays the role of the order parameter of the YMC. The index $a$ runs over the gauge group, for example, $a=1,2,3$ for $SU(2)$, $a=1,2,3,...,8$ for $SU(3)$. Let us consider the ideal scenario where the only component in the Universe is YMC, which is minimally coupled to gravity. The effective action becomes,
\begin{equation}
    \mathcal{S}_{\text{YMC}}=\int d^{4}x\sqrt{-g}\left[\frac{\mathcal{R}}{16\pi G}+\frac{b}{2}F\left(\ln|\frac{F}{\kappa^{2}}|-1\right)\right].
\end{equation}

On applying the variational principle on the action with respect to the metric $g^{\mu\nu}$, we obtain the Einstein equation $G_{\mu\nu}=8\pi GT_{\mu\nu}$. The energy-stress-momentum tensor can be written as,
\begin{equation}\label{Stress-Energy-YM}
T_{\mu\nu}=\sum_{a=1}^{3}\frac{g_{\mu\nu}}{4g^{2}}F^{a}_{\sigma\delta}F^{a\sigma\delta}+\epsilon F^{a}_{\mu\nu}F^{a\sigma}_{\nu},
\end{equation}

Herewith, we assume that the gauge fields are only the functions of time\footnote{A broad discussion on suitable ansatz for the gauge fields can be found in \citep{PVMoniz_1991,PVMoniz_1993}.}. 
They can be expressed as $A_{\mu}=\frac{i}{2}\sigma_{a}A^{a}_{\mu}(t)$, where $\sigma_{a}$ are Pauli's matrices. The time component and the spatial component can be written as $A_{0}=0$ and $A^{a}_{i}=\delta^{a}_{i}(t)$, respectively. The Yang-Mills field tensor is defined as usual,
\begin{equation}\label{Field Strength Tensor}
F^{a}_{\mu\nu}=\partial_{\mu}A^{a}_{\nu}-\partial_{\nu}A^{a}_{\mu}+gf^{abc}A^{b}_{\mu}A^{c}_{\nu},
\end{equation}
where $f^{abc}$ is the structure constant of the group.
Here, the dielectric constant is defined by $\epsilon=2\partial\mathcal{L}_{\text{YMC}}/\partial F$. We get,
\begin{equation}\label{dimensionless}
    y=\frac{\epsilon}{b}=\ln|\frac{F}{\kappa^{2}}|.
\end{equation}

Now, let us find the corresponding energy density and pressure of the energy-stress-momentum tensor in Eq. (\ref{Stress-Energy-YM}). We obtain,
$$\rho_{\text{YMC}}=\frac{1}{2}\epsilon(E^{2}+B^{2})+\frac{1}{2}b(E^{2}-B^{2})\quad\text{and}\quad p_{\text{YMC}}=\frac{1}{6}\epsilon(E^{2}+B^{2})-\frac{1}{2}b(E^{2}-B^{2}).$$

Since, for an expanding Universe, the magnetic component of the Yang-Mills field falls off much faster, and the field effectively becomes the electric type as suggested in \citep{ZHAO_2007}. This consideration simplifies the expressions of energy density and pressure in the following way,
$$\rho_{\text{YMC}}=\frac{E^{2}}{2}(\epsilon+b)\quad\text{and}\quad p_{\text{YMC}}=\frac{E^{2}}{2}\left(\frac{\epsilon}{3}-b\right).$$

We find it convenient to simplify the expressions further by introducing a dimensionless quantity, which we established earlier in Eq. (\ref{dimensionless}). Thereby, further modifying the expressions into the following form,
$$\rho_{\text{YMC}}=\frac{1}{2}b\kappa^{2}(y+1)\exp{(y)}\quad\text{and}\quad p_{\text{YMC}}=\frac{1}{6}b\kappa^{2}(y-3)\exp{(y)}.$$

The EoS of Yang-Mills Condensate becomes,
\begin{equation}
    \omega_{\text{YMC}}=\frac{y-3}{3y+3}.
\end{equation}

Now, finally establishing the correspondence with the FHDE model in the following manner,
\begin{equation}
    \frac{y-3}{3y+3}= -1+\frac{(3\alpha-2)\left(1-\Omega_{\text{de}}\right)}{2\alpha-\Omega_{\text{de}}(3\alpha-2)}.
\end{equation}

On solving, we obtain,
\begin{equation}\label{YMC-field}
E^{2}=\kappa^{2}\exp{\left(3-\frac{12(\alpha-2)}{\alpha+3\alpha\Omega_{\text{de}}-2(3+\Omega_{\text{de}})}\right)}.
\end{equation}

\begin{figure}[H]
\centering
\begin{subfigure}{.5\textwidth}
  \centering
\includegraphics[width=0.9\linewidth]{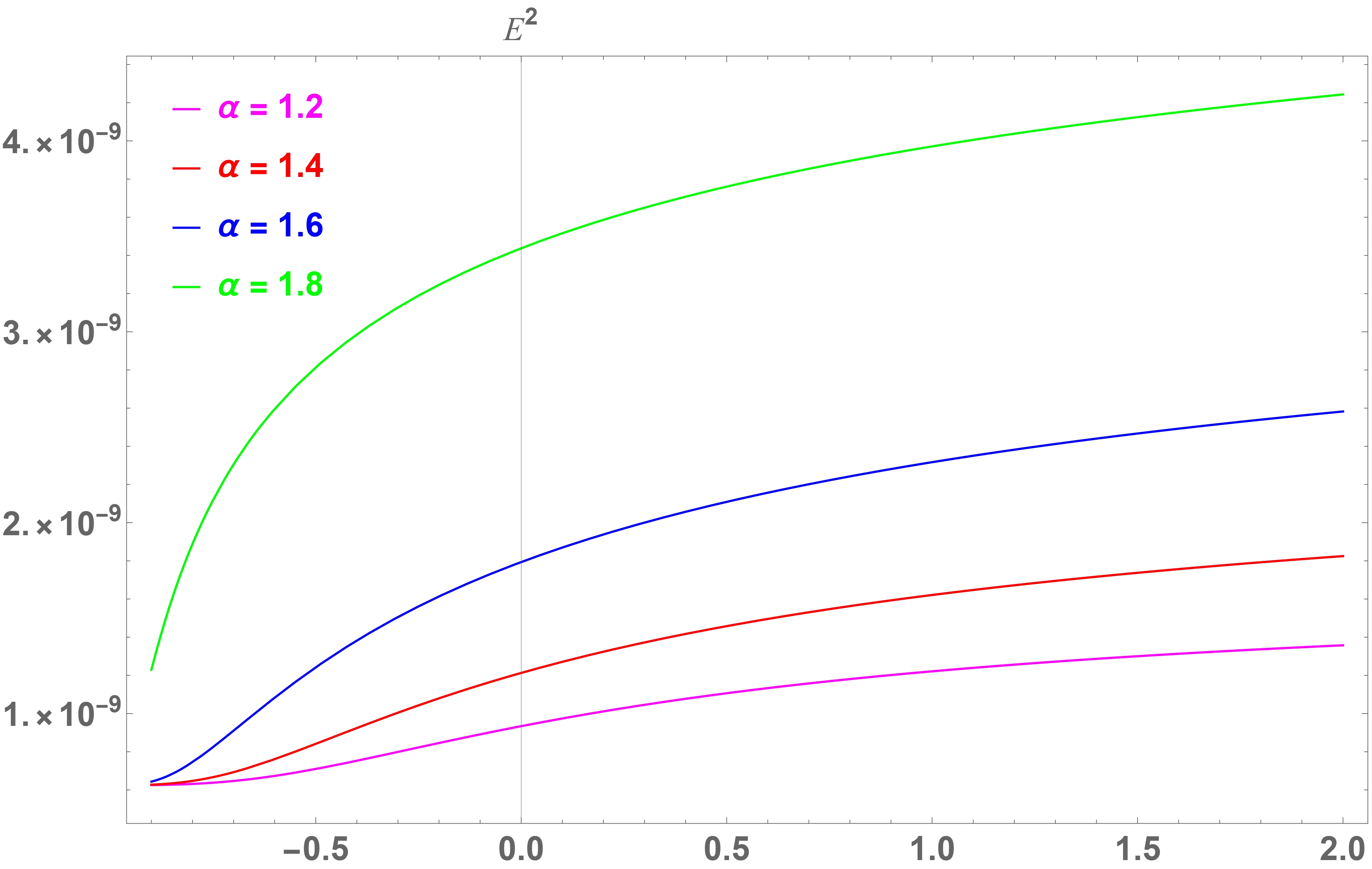}
  \caption{Plot of $E^{2}$ against redshift $z$.}
  \label{Figure 7: (a)}
\end{subfigure}%
\begin{subfigure}{.5\textwidth}
  \centering
\includegraphics[width=0.9\linewidth]{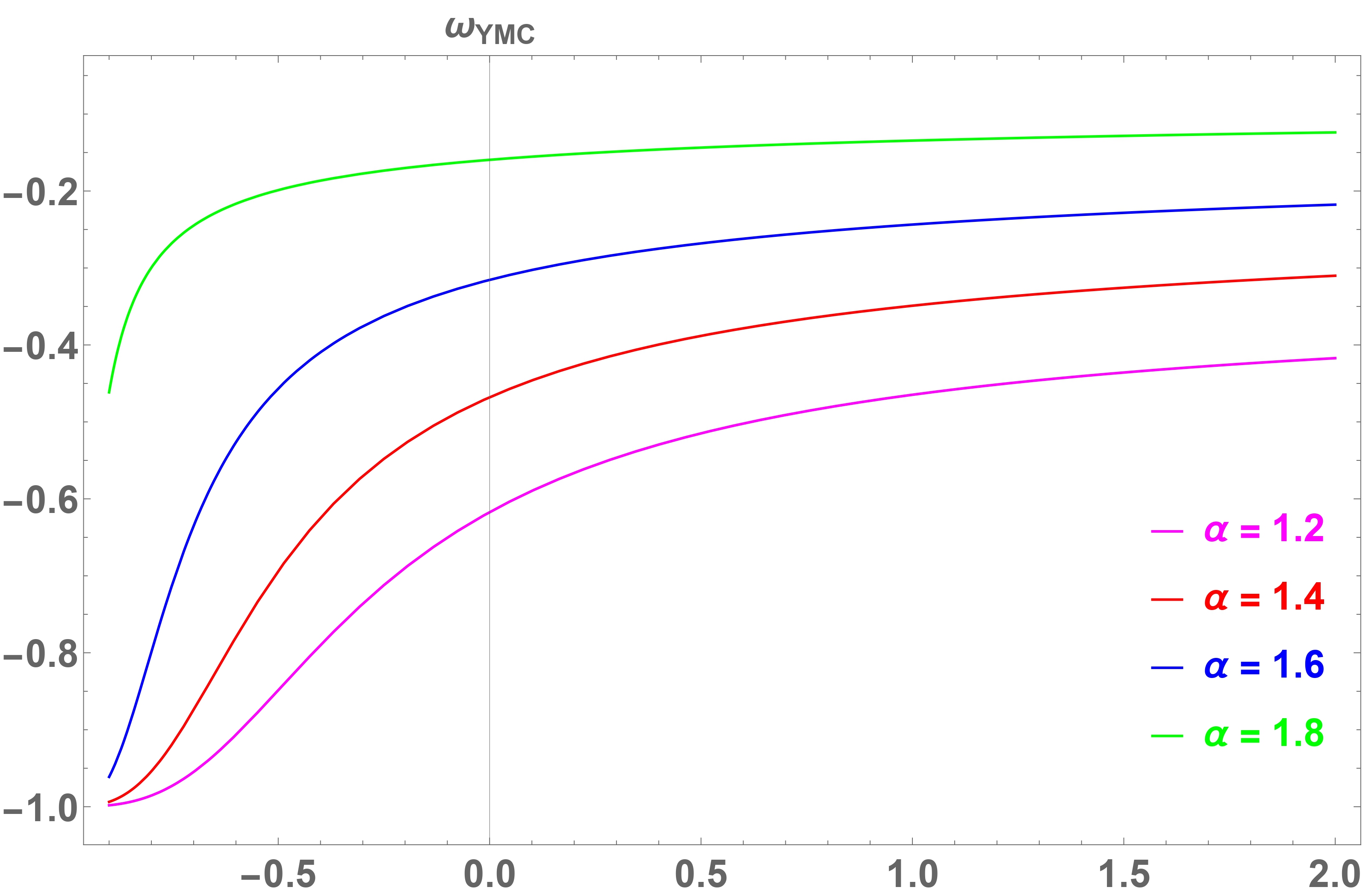}
  \caption{Plot of $\omega_{\text{YMC}}$ against redshift $z$.}
  \label{Figure 7: (b)}
\end{subfigure}
\caption{Plot for $E^{2}$ and $\omega_{\text{YMC}}$ against redshift $z$ for $\alpha=1.2, 1.4, 1.6$ and $1.8$.}
\label{Figure 7}
\end{figure}

Eq. (\ref{YMC-field}) describes the late-time cosmic evolution of the YMC field, which ultimately takes the form of an electric field due to the approximations taken into account to address the late-time Universe as a function of the redshift parameter. The YMC field configuration is particularly different from the rest of the effective field candidates due to the spin-1 gauge fields governing it in a cosmological setting. We plot the evolution of the electric field as shown in Figure \ref{Figure 7: (a)}. Furthermore, we plot the EoS parameter in Figure \ref{Figure 7: (b)}.

In particular:
\begin{itemize}
    \item In Figure \ref{Figure 7: (a)}, we observe that the electric component, $E^{2}$, of the YMC field in fractional holographic reconstruction, decays asymptotically to approach $E^{2}\rightarrow0$ as the redshift $z$ decreases in the far-future limit $z\rightarrow-1$. 

    Furthermore, for larger values of $\alpha$ (e.g., $\alpha=1.8$ and $1.6$), $E^{2}$ shows a highly dynamic evolution, whereas the evolution is less dynamic for smaller values of $\alpha$ (e.g., $\alpha=1.4$ and $1.2$) in the redshift region $-1<z\leq2$. 
    
    \item On the other hand, in Figure \ref{Figure 7: (b)}, we plotted for the EoS parameter, $\omega_{\text{YMC}}\rightarrow-1$ and is possible to identify that as fractional features start to dominate, it conveys a scenario where the YMC field behaves similarly to the $\Lambda$CDM model during dark energy dominated phase.
    
    We remark a similar behaviour for the EoS parameter for quintessence (see Figure \ref{Figure 2}), K-essence (see Figure \ref{Figure 4}) and Dilaton (see Figure \ref{Figure 6}).
\end{itemize}

Thus, Figure \ref{Figure 7: (a)} suggests that higher values of $\alpha$ may not be suitable because they do not ensure a slow evolving field, making it inadequate for 
a nearly constant dark energy.

\subsection{DBI-essence Field}

Recent works in String Theory propose intriguing scenarios where the inflaton, central to inflationary cosmology, is identified with the separation between two branes navigating extra dimensions within a warped throat. This interpretation arises naturally from the action of the system, which is proportional to the volume swept out by the brane as it moves. The volume, determined by the square root of the induced metric, inherently generates a Dirac$-$Born$-$Infeld (DBI) kinetic term, linking this framework directly to the inflationary phase of rapid expansion in the early Universe (see \citep{Ahn_2009} for more details). Let us herewith  consider the DBI scalar field as the dark energy whose action reads as:
\begin{equation}
    \mathcal{S}_{\text{DBI}}=-\int d^{4}x\hspace{1mm}a^{3}(t)\left[T(\varphi)\sqrt
    {1-\frac{\dot{\varphi}^{2}}{T(\varphi)}}+V(\varphi)-T(\varphi)\right],
\end{equation}
where $T(\varphi)=n\dot{\varphi}^{2}$ represents the warped brane 
tension, but we will refer to it as the kinetic energy $X_{\text{DBI}}$, and $V(\varphi)$ is the potential arising from
interactions with the Ramond-Ramond fluxes (see \citep{Tong}). We then proceed to obtain the energy density and pressure expressions as:
   $$\rho_{\text{DBI}}=\left(\eta-1\right)T(\varphi)+V(\varphi)\quad \text{and}\quad p_{\text{DBI}}=\left(\frac{\eta-1}{\eta}\right)T(\varphi)-V(\varphi).$$

Here, $\eta$ represents the Lorentz Boost factor, which can be written as: 
\begin{equation}
    \eta=\sqrt{\frac{1}{1-\dot{\varphi}^{2}/T(\varphi
    )}}.
\end{equation}

Moreover, the EoS parameter for DBI--essence dark energy can be written as:
\begin{equation}
    \omega_{\text{DBI}}=\frac{(\eta-1)T(\varphi)-\eta V(\varphi)}{\eta(\eta-1)T(\varphi)+\eta V(\varphi)}.
\end{equation}

Now, we are establishing the correspondence with the FHDE model. We obtain:
\begin{equation}
    \frac{(\eta-1)T(\varphi)-\eta V(\varphi)}{\eta(\eta-1)T(\varphi)+\eta V(\varphi)}=-1+\frac{(3\alpha-2)\left(1-\Omega_{\text{de}}\right)}{2\alpha-\Omega_{\text{de}}(3\alpha-2)}.
\end{equation}

Upon solving the above-written correspondence equation, we obtain the expressions for $X_{\text{DBI}}$ and $V_{\text{DBI}(\varphi)}$ as:
\begin{equation}\label{DBI-field}
X_{\text{DBI}}=n\dot{\varphi}^{2}_{\text{DBI}}=\frac{3 (3 \alpha -2) H^2 (n-1) \sqrt{\frac{n}{n-1}} (1-\Omega_{\text{de}} ) \Omega_{\text{de}} }{2 \alpha -(3 \alpha -2) \Omega_{\text{de}} },
\end{equation}

\begin{equation}\label{DBI-potential}
    V_{\text{DBI}}(\varphi)=3 H^2 \Omega_{\text{de}} -\frac{3 (3 \alpha -2) H^2 (n-1) \sqrt{\frac{n}{n-1}} \left(\sqrt{\frac{n}{n-1}}-1\right) (1-\Omega_{\text{de}} ) \Omega_{\text{de}} }{2 \alpha -(3 \alpha -2) \Omega_{\text{de}} }.
\end{equation}

\begin{figure}[H]
\centering
\begin{subfigure}{.5\textwidth}
  \centering
\includegraphics[width=0.9\linewidth]{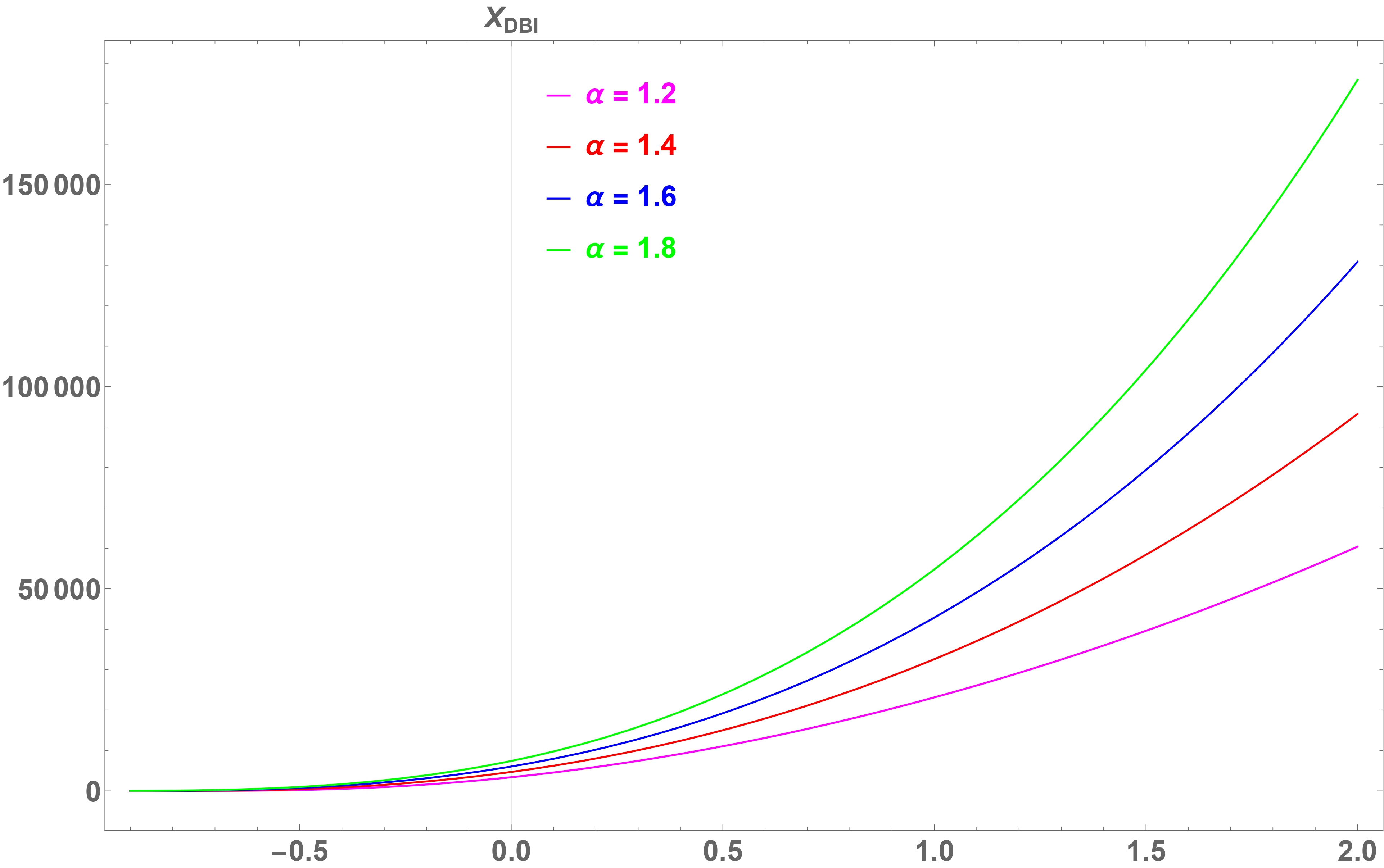}
 \caption{Plot of $X_{\text{DBI}}$ against redshift $z$.}
  \label{Figure 8: (a)}
\end{subfigure}%
\begin{subfigure}{.5\textwidth}
  \centering
  \includegraphics[width=0.9\linewidth]{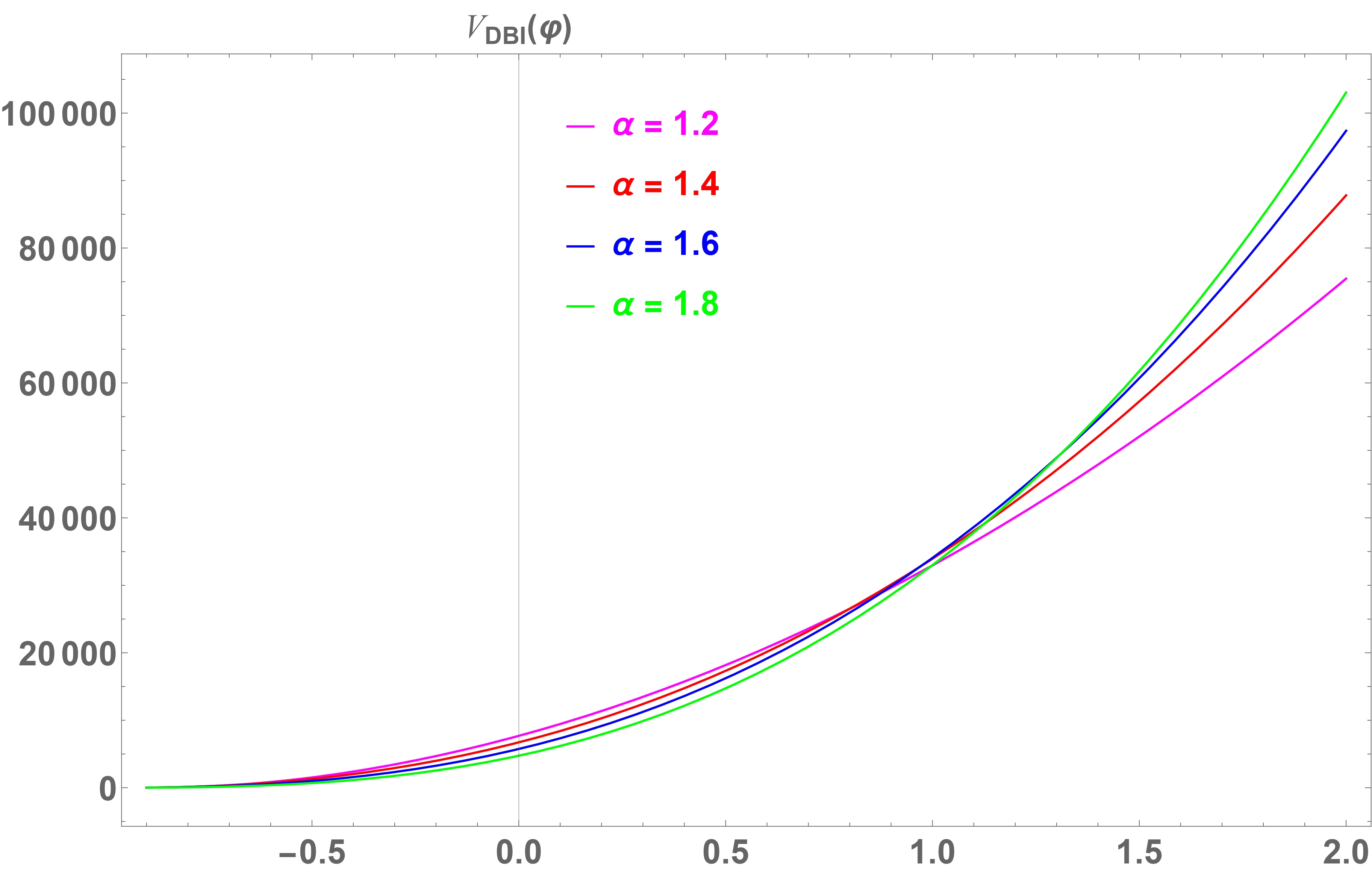}
  \caption{Plot of $V_{\text{DBI}}(\varphi)$ against redshift $z$.}
  \label{Figure 8: (b)}
\end{subfigure}
\caption{Plot of $X_{\text{DBI}}$ and $V_{\text{DBI}}(\varphi)$ against redshift $z$ for $\alpha=1.2, 1.4, 1.6$ and $1.8$. Here $n=1.5$.}
\label{Figure 8}
\end{figure}

Eq. (\ref{DBI-field}) and Eq. (\ref{DBI-potential}) describe the late-time cosmic evolution of the kinetic energy, $X_{\text{DBI}}(\varphi)$, and potential, $V_{\text{DBI}}(\varphi)$, in correspondence with the FHDE model. The evolution is illustrated in Figure \ref{Figure 7} and Figure \ref{Figure 8}. 

\begin{figure}[H]
\centering  \includegraphics[width=0.5\linewidth]{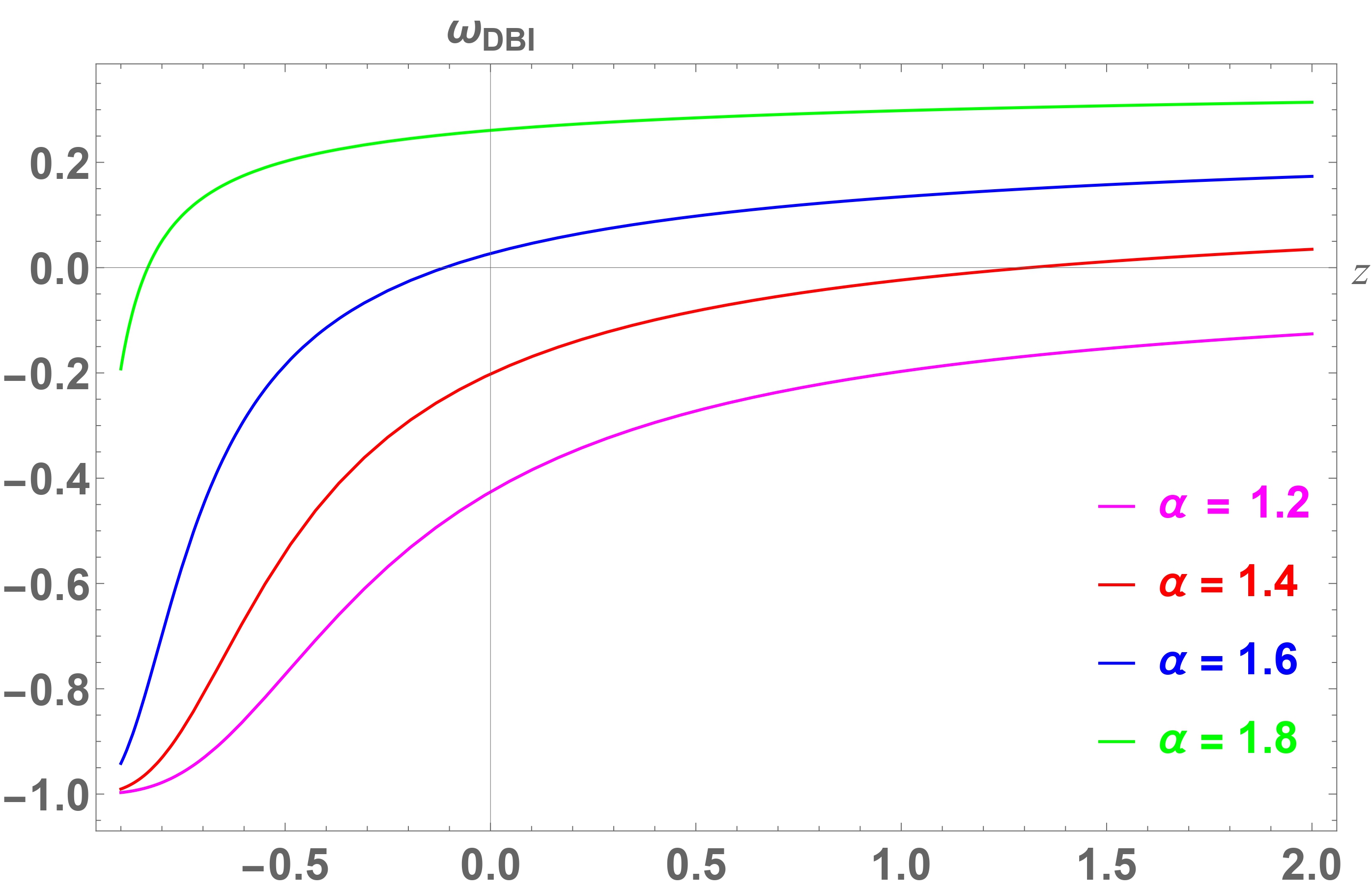}
\caption{Plot of $\omega_{\text{DBI}}$ against redshift $z$ for $\alpha=1.2, 1.4, 1.6$ and $1.8$; here $c=0.01$.}
\label{Figure 9}
\end{figure}

We notice the following:
\begin{itemize}

    \item In Figure \ref{Figure 8: (a)} and Figure \ref{Figure 8: (b)}, we notice that the kinetic energy, $X_{\text{DBI}}$, and the potential energy, $V_{\text{DBI}}(\varphi)$, of the DBI-essence field in fractional holographic reconstruction, decays to asymptotically approach zero, respectively, as the redshift $z$ decreases in the far-future limit $z\rightarrow-1$. 

In particular:

\begin{itemize}
    \item     This behaviour of $X_{\text{DBI}}$ and $V_{\text{DBI}}(\varphi)$ is consistent across all values of the fractional parameter $\alpha$.
    
    \item For larger values of $\alpha$ (e.g., $\alpha=1.8$ and $1.6$), the evolution for $X_{\text{DBI}}$ shows a highly dynamic evolution, whereas the evolution is less dynamic for smaller values of $\alpha$ (e.g., $\alpha=1.4$ and $1.2$). 
    
    \item In Figure \ref{Figure 8: (b)}, it is noticed a similar behaviour to what we observed in quintessence (see Figure \ref{Figure 1: (a)}), however, for a different redshift region: $-1<z\leq0.99$; where the $V_{\text{DBI}}(\varphi)$ decreases less dynamically for larger values of $\alpha$ (e.g., $\alpha=1.8$ and $1.6$). 

\end{itemize}
    
    \item In Figure \ref{Figure 9}, we plotted for the EoS parameter, $\omega_{\text{DBI}}\rightarrow-1$ and is possible to identify that as fractional features start to dominate,  it represents a scenario where the DBI-essence field behaves similarly to the $\Lambda$CDM model during dark energy dominated phase. Moreover, the EoS parameter takes a different value at $z=0$ from what we observed in Quintessence, K-essence, Dilaton and YMC fields.
\end{itemize}

In summary regarding  this subsection, Figure \ref{Figure 8: (a)} suggests that a higher value of $\alpha$ may not provide a suitable scenario because it does not ensure a slow-evolving field, 
not mimicking 
a nearly constant dark energy behaviour. On the other hand, in Figure \ref{Figure 8: (b)}, a higher value of $\alpha$ is preferable (over smaller values) because it induces a suitable situation for the potential in the redshift region $-1<z\leq0.99$.

\subsection{Tachyonic Field}

The tachyon field is a possible candidate for inflation and dark energy at high energies as suggested in \citep{Mazumdar_2001, Feinstein_2002, Piao_2002, gibbons2002cosmological}. However, the nature of the potential for self-interaction, $V(\varphi)$, acts as a deciding factor to ultimately obtain the desired dark energy behaviour. By setting the correspondence with FHDE, we attempt to express the potential explicitly, which will help us later study the tachyon field's cosmological evolution. We begin with the effective tachyon scalar field Lagrangian,
\begin{equation}\label{T-Lag}
\mathcal{L}_{\text{t}}=-V_{\text{t}}\left(\varphi\right)\sqrt{1-g^{\mu\nu}\partial_{\mu}\varphi\partial_{\nu}\varphi}
\end{equation}

Its corresponding energy-stress-momentum tensor can be written similarly to that of a perfect-fluid tensor in the following way,
\begin{equation}
    T_{\mu\nu}^{\varphi}=(p_{\text{t}}+\rho_{\text{t}})u_{\mu}u_{\nu}+p_{\text{t}}g_{\mu\nu}
\end{equation}

The energy density and pressure can be written as,

\begin{equation}\label{pressureanddensity}
    p_{\text{t}}(t)=-V_{\text{t}}\left(\varphi\right)\sqrt{1-\dot{\varphi}_{\text{t}}^{2}}\quad\text{and}\quad\rho_{\text{t}}(t)=\frac{V_{\text{t}}\left(\varphi\right)}{\sqrt{1-\dot{\varphi}_{\text{t}}^{2}}}.
\end{equation}

Moreover, the EoS parameter can be expressed as,

\begin{equation}\label{EoS-T}
\omega_{\text{t}}=\dot{\varphi}^{2}_{t}-1.
\end{equation}

Now, establishing the correspondence with FHDE provides us with
\begin{equation}
    \dot{\varphi}^{2}_{\text{t}}-1=-1+\frac{(3\alpha-2)\left(1-\Omega_{\text{de}}\right)}{2\alpha-\Omega_{\text{de}}(3\alpha-2)}.
\end{equation}

Upon solving the above-written correspondence equation, we obtain the expression for $\dot{\varphi}_{\text{t}}$ and $V_{\text{t}}(\varphi)$ as,
\begin{equation}\label{t-field}
X_{\text{t}}=\frac{\Dot{\varphi}^{2}_{\text{t}}}{2}=\frac{\left(3\alpha-2\right)\left(1-\Omega_{\text{de}}\right)}{4\alpha-\Omega_{\text{de}}\left(6\alpha-4\right)},
\end{equation}
\begin{equation}\label{t-potential}
V_{\text{t}}\left(z\right)=3H^{2}\Omega_{\text{de}}\cdot\left[1-\frac{\left(3\alpha-2\right)\left(1-\Omega_{\text{de}}\right)}{2\alpha-\Omega_{\text{de}}\left(3\alpha-2\right)}\right]^{\frac{1}{2}}.
\end{equation}
Let us now elaborate on this setting, starting with the following plots.

\begin{figure}[H]
\centering
\begin{subfigure}{.5\textwidth}
  \centering
\includegraphics[width=0.9\linewidth]{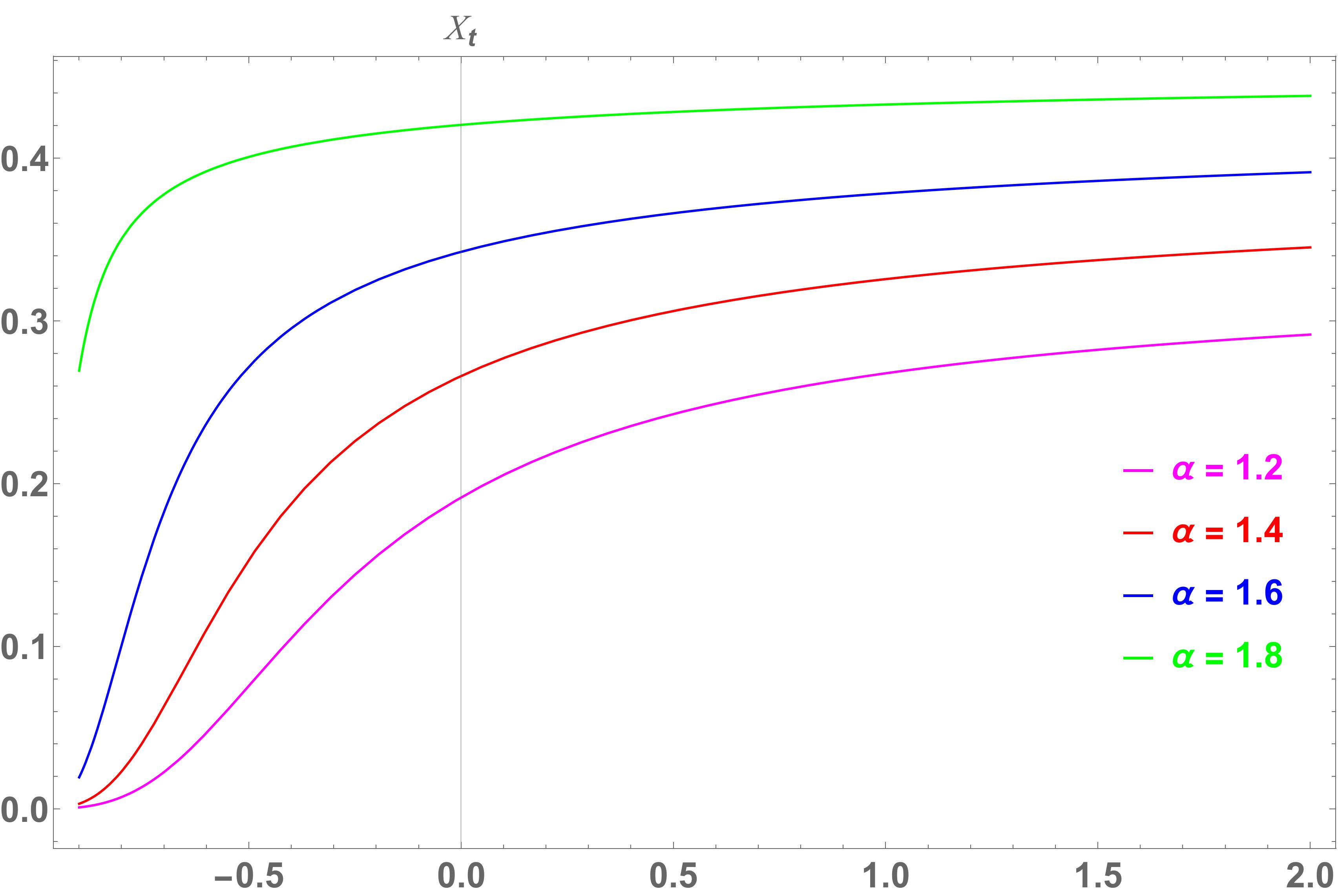}
  \caption{Plot of $X_{\text{t}}$ against redshift $z$.}
  \label{Figure 10: (a)}
\end{subfigure}%
\begin{subfigure}{.5\textwidth}
  \centering
  \includegraphics[width=0.9\linewidth]{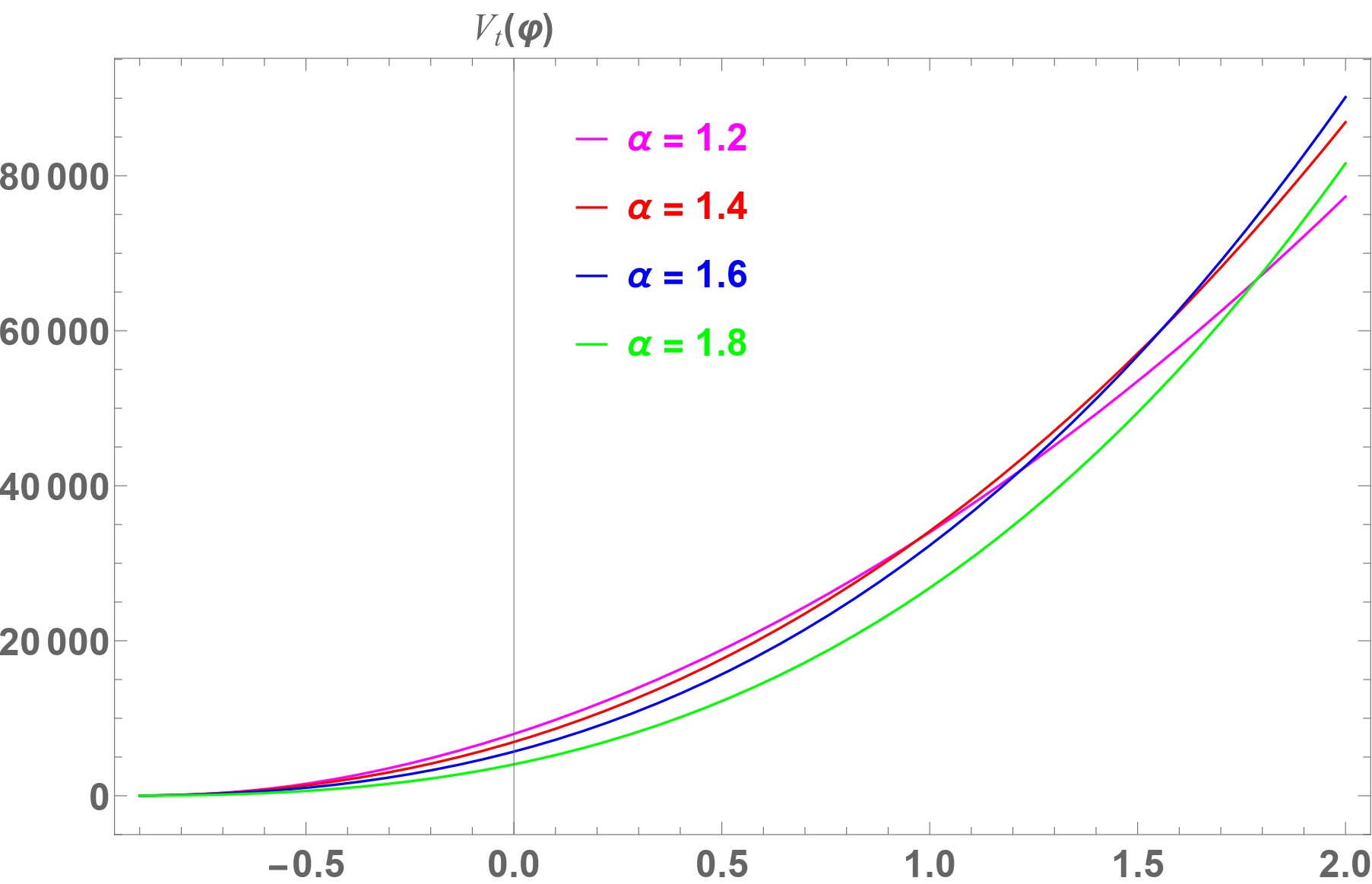}
  \caption{Plot of $V_{\text{t}}(\varphi)$ against redshift $z$.}
  \label{Figure 10: (b)}
\end{subfigure}
\caption{Plot of $X_{\text{t}}$ and $V_{\text{t}}(\varphi)$ against redshift $z$ for $\alpha=1.2, 1.4, 1.6$ and $1.8$.}
\label{Figure 10}
\end{figure}

\begin{figure}[H]
\centering  \includegraphics[width=0.5\linewidth]{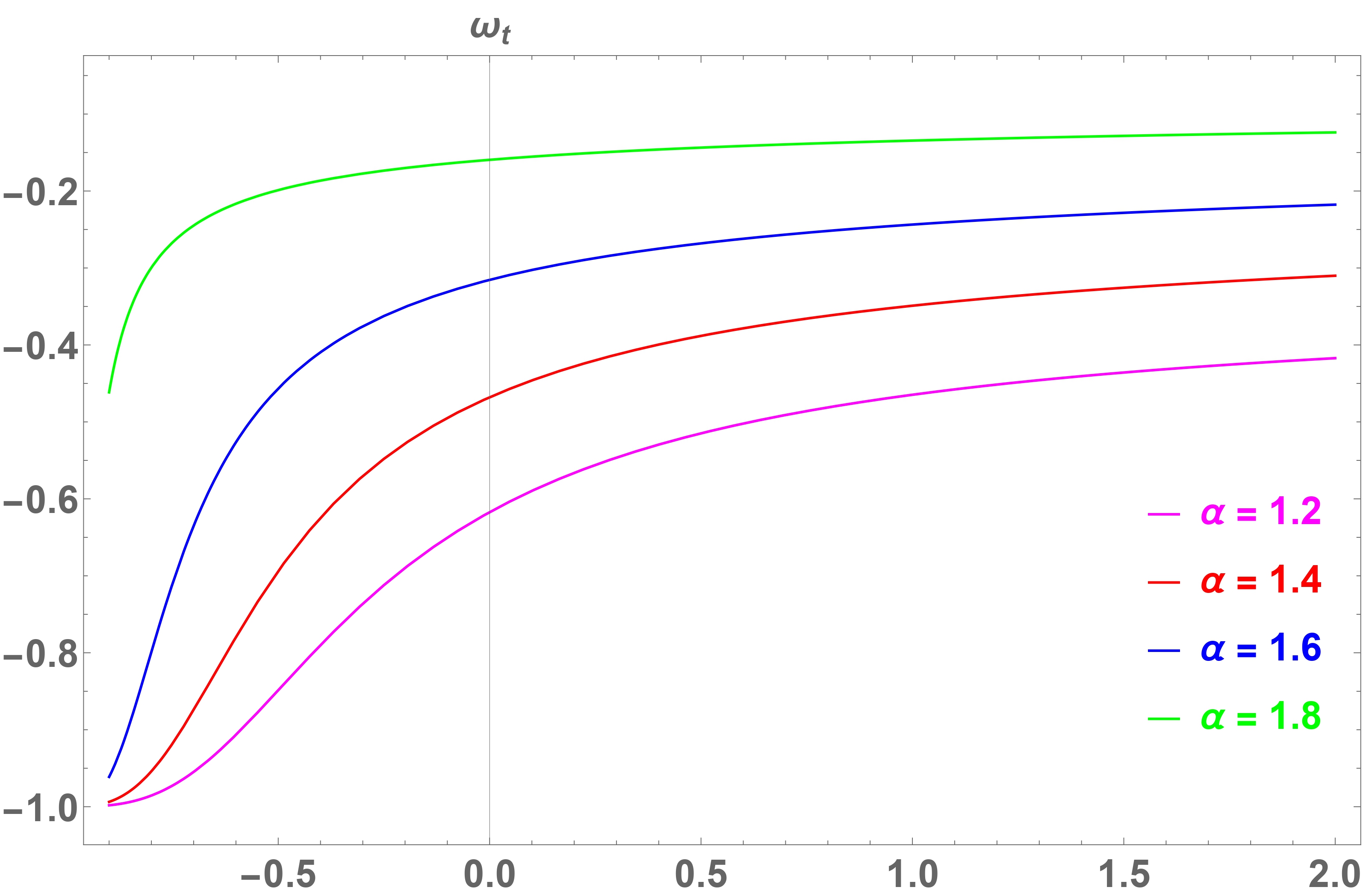}
\caption{Plot of $\omega_{\text{t}}$ against redshift $z$ for $\alpha=1.2, 1.4, 1.6$ and $1.8$; here $c=0.01$.}
\label{Figure 11}
\end{figure}

We notice the following:
\begin{itemize}
    \item In Figure \ref{Figure 10: (a)} and Figure \ref{Figure 10: (b)}, we perceive that the kinetic energy, $X_{\text{t}}$, and the exponential function, $V_{\text{t}}(\varphi
)$, of the Tachyon field in fractional holographic reconstruction, decays to asymptotically approach $X_{\text{t}}\rightarrow0$, and $V_{\text{t}}(\varphi
)\rightarrow0$, respectively, as the redshift $z$ decreases in far-future limit $z\rightarrow-1$.

With detail:

\begin{itemize}
    \item For larger values of $\alpha$ (e.g., $\alpha=1.8$ and $1.6$), $X_{\text{t}}$ shows a highly dynamic (or faster) evolution, whereas the evolution is less dynamic (or slower) for smaller values of $\alpha$  (e.g., $\alpha=1.4$ and $1.2$) in the redshift region $-1\leq z\leq 2$.

\item On the other hand, in Figure \ref{Figure 10: (b)}, we observe that for larger values of $\alpha$ (e.g., $\alpha=1.8$ and $1.6$), the potential evolution is less dynamical (or slower) and in contrast the evolution for smaller values of $\alpha$ (e.g., $\alpha=1.2$ and $1.4$), is highly dynamic in the redshift region $-1<z\leq1.8$.

\end{itemize}

\item In Figure \ref{Figure 11}, we plotted for the EoS parameter, $\omega_{\text{t}}$, of the fractional holographic reconstructed Tachyon model. We can retrieve that a suitable evolution of $\omega_{\text{t}}\rightarrow-1$ is possible to identify as fractional features start to dominate because it represents a scenario where the Tachyon scalar field behaves similarly to the $\Lambda$CDM model during dark energy dominated phase. 

We discern a similar behaviour for the EoS parameter for quintessence (see Figure \ref{Figure 2}), K-essence (see Figure \ref{Figure 4}) and Yang-Mills condensate (see Figure \ref{Figure 7: (b)}) in the far-future limit.
\end{itemize}

Briefly, Figure \ref{Figure 10: (a)} suggests that a higher value of $\alpha$ may not provide a suitable scenario because it does not ensure a slow-evolving field, 
not suggesting
a nearly constant dark energy behaviour. On the other hand, in Figure \ref{Figure 10: (b)}, a higher value of $\alpha$ is preferable (over smaller values) because it induces a suitable situation for the potential in the redshift region $-1<z\leq1.8$.

\section{Discussion and Outlook}\label{conc}





The generic purpose embraced  in this paper was to 
 contribute towards a  research programme that promotes 
 an interplay between
 the Holographic Principle and particular features from Fractional Calculus \citep{FC, Miller1993AnIT, Grigoletto2013FractionalVO}. 
This setting is  Fractional Holographic Dark Energy (FHDE) and was introduced in \citep{Trivedi:2024inb}, supported with concrete examples of applications.

The concrete focus we took and that guided our work 
throughout was to make progress, investigating  whether FHDE  can be a realistic and valued pretender to explain the late-time accelerated expansion of the universe. To be more precise, we methodically selected and appraised suitable field frameworks, such that they could  satisfactorily  encompass  the alluring dynamic properties of FHDE as explained in \citep{Trivedi:2024inb}.

In more detail, we have considered  
the late time 
evolution of the Universe within a twofold strategy. On the one hand, we employed 
particular effective field 
configurations, importing features from known settings involving spin-$0$ and spin-$1$ features. On the other hand, we used  particular fractional calculus elements \citep{ortigueira2011fractional,ortigueira2017fractional,ortigueira2012relation,ortigueira2023fractional,valerio2023variable,bengochea2023operational,ortigueira2024factory} 
that have been considered in several gravitational scenarios \citep{Garcia-Aspeitia:2022uxz, Micolta-Riascos:2023mqo, LeonTorres:2023ehd, Calcagni:2009kc}.

A twofold query was discussed. On the one hand, can  FHDE \citep{Trivedi:2024inb}  indeed be perceived as a promising aspirant with the assistance of suitable field frameworks? On the other hand, can a suitable correspondence between those frameworks be established with the FHDE model, unveiling scenarios for dynamic dark energy from a new angle? Assuming  the Hubble horizon as the IR cut-off,  we  investigated  the following 
effective field configurations: ($i$) Quintessence, ($ii$) Kinetic Quintessence (K-essence), ($iii$) Dilaton, ($iv$)Yang-Mills condensate, ($v$) Dirac-Born-Infeld-essence (DBI-essence), and ($vi$) Tachyon scalar field. To prospect whether such  correspondence for  the above-mentioned ($i$)-($vi$) configurations 
can be found 
within the FHDE scenario, 
%
we reconstructed two quantities: (a) kinetic energy, $X_{i}$, and (b) potential, $V_{i}(\varphi)$, with varying redshift, where $i$ is a dummy index representing the effective field models. 
We then produced suitable plots of these quantities and the EoS parameter for various values of the fractional parameter $\alpha$ such as $\alpha=1.2,1.4,1.6$ and $1.8$.








Our results, concerning the reconstruction of FHDE features by means of selected field configurations, can be assembled as follows:

\begin{itemize}
    \item Let us firstly review  the behaviour of  $X_{i}$:
    \begin{itemize}
        \item Among all the field 
configurations we used, a common feature
is remarked for Quintessence (Figure \ref{Figure 1: (a)}), K-essence (Figure \ref{Figure 3: (a)}), DBI-essence (Figure \ref{Figure 8: (a)}), and Tachyon scalar field (Figure  \ref{Figure 10: (a)}) 
which establishes 
that over the redshift region 
$-1<z\leq2$, 
an adequate course for  $X_{i}$ is appropriately 
described by a \textit{smaller} value of the fractional parameter, for example, $\alpha=1.2$. Concretely, it bears a less dynamic 
or slowly evolving behaviour as the redshift $z$ decreases. This behaviour for smaller $\alpha$ values strongly suggest that field configuration exhibits slow variation, associated with the field energy asymptotically approaching stabilisation 
with the 
decreasing 
of the 
field potential $V_{i}(\varphi)$ (for all values of $\alpha$) as $z\rightarrow-1$ during dark energy domination. 

\item In the case of Dilaton (Figure \ref{Figure 5: (a)}), the \textit{larger} values of $\alpha$, like $\alpha=1.8$ and $1.6$, show a less dynamic behaviour as the redshift decreases below $z<1.4$, making larger values of $\alpha$ more suitable in the redshift region $-1<z\leq1.5$. 

This distinguishes Dilaton as an exception compared to other effective field configurations, wherein a smaller value of $\alpha$ typically indicates suitable behaviour as the Universe transitions to dark energy domination in the asymptotic future limit $z \rightarrow -1$. 

\item Moreover, for gauge field configuration, i.e.,  YMC, we notice  that the electric field component, $E^{2}$, represents 
a scenario where the evolution described by 
\textit{smaller} values of $\alpha$ 
is more suitable compared to evolution described by larger values of $\alpha$ as redshift $z$ decreases.

\end{itemize}

\item Regarding 
the pattern of $V_{i}(\varphi)$ and the coupling function $f_{\text{kq}}(\varphi)$:

\begin{itemize}
    
\item For Quintessence (Figure \ref{Figure 1: (b)}), DBI-essence (Figure \ref{Figure 8: (b)}), and Tachyon (Figure \ref{Figure 10: (b)}) and the coupling function $f_{\text{kq}}(\varphi)$ of K-essence (Figure \ref{Figure 3: (b)}), they exhibit a common behaviour. 

In these models, the potential decreases less dynamically for \textit{larger} values of $\alpha$, such as $\alpha = 1.8$ and $\alpha = 1.6$, as the redshift $z$ decreases below certain thresholds. 

Specifically, for Quintessence, this less dynamic decrease occurs as the redshift decreases below $z < 0.55$ (Figure \ref{Figure 1: (b)}), for DBI-essence below $z < 0.99$ (Figure \ref{Figure 8: (b)}), and for Tachyon below $z < 1.8$ (Figure \ref{Figure 10: (b)}). 

Conversely, the coupling function $f_{\text{kq}}(\varphi)$ for K-essence demonstrates a less dynamic evolution for $\alpha = 1.8$ across the entire redshift range $-1 < z \leq 2$, without exhibiting a switch in $\alpha$ value, unlike the observations in Quintessence, DBI-essence, and Tachyon. 

\item For Dilaton (Figure \ref{Figure 5: (b)}), it is observed that the potential $\beta\exp{(\lambda\varphi)}X_{\text{d}}$ exhibits a less dynamic behaviour for \textit{smaller} values of $\alpha$, such as $\alpha = 1.2$ and $\alpha = 1.4$, in contrast to Quintessence, DBI-essence, Tachyon, and K-essence. The potential, $\beta\exp{(\lambda\varphi)}X_{\text{d}}$, asymptotically approaches $\Lambda$CDM behaviour in the far-future limit $z \rightarrow -1$. 

\end{itemize}


\item Regarding the EoS parameter, the following should be mentioned:

\begin{itemize}
    \item In the case of Quintessence (Figure \ref{Figure 2}), K-essence (Figure \ref{Figure 4}), Dilaton (Figure \ref{Figure 6}), YMC (Figure \ref{Figure 7: (b)}), and Tachyon (Figure \ref{Figure 11}), it can be noticed that for \textit{small} values of $\alpha$, like $\alpha=1.2$ and $1.4$, the EoS parameter for these effective field configurations asymptotically approaches $\Lambda$CDM behaviour: $\omega_{i}(z)\rightarrow-1$; in the far-future limit $z\rightarrow-1$. 
    
    In particular, for $\alpha=1.2$, the EoS parameter takes a value at $z=0$, which is in close agreement with the recent constraint on the EoS parameter by DESI Collaboration \citep{desicollaboration2024desi2024vicosmological, adame2024desi1,adame2024DESI}. 

    \item However, for DBI-essence (Figure \ref{Figure 9}), the EoS parameter at $z = 0$ deviates slightly, being 
    \textit{larger} than observed in other effective field configurations. 

\end{itemize}

\end{itemize}





In closing, we successfully reconstructed selected field (kinetic and potential) features within the fractional holographic framework introduced in \citep{Trivedi:2024inb}. Our analysis demonstrated that fractional modifications help to alleviate quantum instabilities by ensuring that the EoS parameter remains above the phantom divide i.e., $\omega(z)<-1$. We found that smaller values of the fractional parameter $\alpha$ lead to a suitable evolution that asymptotically approaches a $\Lambda$CDM behaviour in the far-future limit $z\rightarrow-1$. In particular, for string theory-inspired field configurations, such as Dilaton, DBI-essence, and the Tachyon field, we found that their EoS parameter exhibits a suitable late-time cosmic acceleration, especially for small values of $\alpha$, prompting that fractional modifications can provide an effective description of dark energy in such frameworks. However, since string theory lacks direct experimental verification, the exact nature of dark energy within these models remains an open question. On the other hand, Quintessence, K-essence, and Yang-Mills condensates (YMC) emerge as strong alternatives, especially when employing fractional modifications. These field configurations provide a viable framework for exploring dark energy while remaining testable against observational constraints such as CMB anisotropies, BAOs, and Supernova data.



Thus, upon the progress displayed in \citep{Trivedi:2024inb} and herewith this paper, we place confidence in the FHDE scenario and list a few subsequent lines that we aim to pick up as future work.  

On the one hand, if fractional calculus bears significance in at least dark energy physics, then it must be "observed". How can we design a test? We are pondering about a behaviour emerging for late-time data that only $\alpha \neq 2$  and no other feature can bring. 

One of the possibilities is to explore the impact of FHDE on the Integrated Sachs-Wolfe (ISW) effect, given its sensitivity to evolving gravitational potentials
in the late universe. 
A fractional calculus-inspired modification of dark energy (e.g., FHDE) might imprint detectable deviations in Cosmic Microwave Background-Large Scale Structure (CMB-LSS) correlations and provide an alternative observational probe beyond the constraints of SNe Ia (see \citep{2016, Giannantonio_2012} and \citep{St_lzner_2018} for further commentary on the ISW effect).

Furthermore, future work could also refine observational constraints to better distinguish these 
effective field configurations reconstructed with fractional holographic considerations. 

Additionally, the dynamic stability of fractional holographic models in a non-flat Universe, i.e., $k\neq0$, plus the interaction between dark matter-dark energy, exploring connections with modified gravity frameworks, and employing more evolved cut-off(s) such as Granda-Oliveros and Nojiri-Odintsov cut-off, all these warrant further investigation within the paradigm of the FHDE scenario. 

Future work could also be done to investigate the emergence of asymptotic future singularities such as "big rip" at the classical level \citep{caldwell2002phantom}.

\section*{Acknowledgements}
PM acknowledges the FCT grant UID-B-MAT/00212/2020 at CMA-UBI plus
the COST Actions CA23130 (Bridging high and low energies in search of
quantum gravity (BridgeQG)) and CA23115 (Relativistic Quantum Information (RQI)). 

\appendix

\section{Appendix}\label{a}
In \citep{Trivedi:2024inb}, we describe the fractional density parameter $\Omega_{\text{de}}$ using a differential equation which can be written in the most general form as,
\begin{equation}\label{Differential Equation}
    \left(1+z\right)\frac{d\Omega_{\text{de}}}{dz}=3\left(\frac{\alpha-2}{3\alpha-2}\right)\left(\frac{\left(3\alpha-2\right)\left(1-\Omega_{\text{de}}\right)+2\alpha\gamma}{2\alpha-\Omega_{\text{de}}\left(3\alpha-2\right)}-\gamma
\right)\Omega_{\text{de}}.
\end{equation}

Here $\gamma$ represents the coupling constant. However, within the scope of this work, we stick with a non-interacting scenario where $\gamma=0$. The Eq. (\ref{Differential Equation}) becomes,
\begin{equation}\label{DE}
\left(1+z\right)\frac{d\Omega_{\text{de}}}{dz}=\frac{3\left(\alpha-2\right)\left(1-\Omega_{\text{de}}\right)}{2\alpha-\Omega_{\text{de}}\left(3\alpha-2\right)}\Omega_{\text{de}}.
\end{equation}

Upon solving this differential equation, we obtain the following expression for $\Omega_{\text{de}}$ as,
\begin{equation}\label{Omega_de}
\Omega_{\text{de}}=\Omega_{0}\left(\frac{1-\Omega_{\text{de}}}{1-\Omega_{0}}\right)^{\frac{2-\alpha}{2\alpha}}\left(1+z\right)^{\frac{3\left(\alpha-2\right)}{2\alpha}}.
\end{equation}

For the sake of completeness, an implicit equation of fractional density parameter for dark matter can be obtained as a function of redshift by using the Friedmann equation: $\Omega_{\text{dm}}=1-\Omega_{\text{de}}$. We obtain,
\begin{equation}
    \Omega_{\text{dm}}=1-\Omega_{0}\left(\frac{1-\Omega_{\text{de}}}{1-\Omega_{0}}\right)^{\frac{2-\alpha}{2\alpha}}\left(1+z\right)^{\frac{3\left(\alpha-2\right)}{2\alpha}}.
\end{equation}

Now we proceed to obtain an implicit expression for the Hubble parameter $H(z)$ as a function of the redshift parameter by plugging in $\Omega_{\text{de}}=c^{2}H^{\frac{\alpha-2}{\alpha}}$ in Eq. (\ref{DE}). We obtain,
\begin{equation}\label{hubble}
H\left(z\right)=H_{0}\left[\frac{c^{2}-H_{0}^{\frac{2-\alpha}{\alpha}}}{c^{2}-H^\frac{2-\alpha}{\alpha}}\right]^{\frac{\alpha}{3\alpha-2}}\left(1+z\right)^{\frac{3\alpha}{(3\alpha-2)}}.
\end{equation}

These expressions play a pivotal role in obtaining the desired plots for the dynamic-dark-energy candidates in correspondence with the FHDE model. We have set the constants  $H_{0}=70\text{km}\cdot\text{Mpc}^{-1}\cdot\text{s}^{-1}$, $\Omega_{\text{0}}=0.69$ and $c=0.01$.

\bibliographystyle{unsrtnat}
\bibliography{references}  






\end{document}